\documentclass[aip,jcp,twocolumn,longbibliography,reprint]{revtex4-2}

\usepackage{amssymb,amsmath}

\usepackage{epsfig}

\usepackage{graphicx}

\usepackage{color}

\newcommand\beq{\begin{equation}}
\newcommand\eeq{\end{equation}}
\newcommand\beqa{\begin{eqnarray}}
\newcommand\eeqa{\end{eqnarray}}
\newcommand{\al}{\alpha}
\newcommand{\dd}{\text{d}}

\usepackage{epstopdf}

\begin{document}
\title{Time-dependent homogeneous states of binary granular suspensions}
\author{Rub\'en G\'omez Gonz\'alez}
\email{ruben@unex.es}
\affiliation{Departamento de
F\'{\i}sica, Universidad de Extremadura, E-06071 Badajoz, Spain}
\author{Vicente Garz\'o}
\email{vicenteg@unex.es} \homepage{http://www.unex.es/eweb/fisteor/vicente/}
\affiliation{Departamento de F\'{\i}sica and Instituto de Computaci\'on Cient\'{\i}fica Avanzada (ICCAEx), Universidad de Extremadura, E-06071 Badajoz, Spain}

\begin{abstract}
The time evolution of a homogeneous bidisperse granular suspension is studied in the context of the Enskog kinetic equation. The influence of the surrounding viscous gas on the solid particles is modeled via a deterministic viscous drag force plus a stochastic Langevin-like term. It is found first that, regardless of the initial conditions, the system reaches (after a transient period lasting a few collisions per particle) a universal unsteady hydrodynamic regime where the distribution function of each species not only depends on the dimensionless velocity (as in the homogeneous cooling state) but also on the instantaneous temperature scaled with respect to the background temperature. To confirm this result, theoretical predictions for the time-dependent partial temperatures are compared against direct simulation Monte Carlo (DSMC) results; the comparison shows an excellent agreement confirming the applicability of hydrodynamics in granular suspensions. Also, in the transient regime, the so-called Mpemba-like effect (namely, when an initially hotter sample cools sooner than the colder one) is analyzed for inelastic collisions. The theoretical analysis of the Mpemba effect is performed for initial states close to and far away from the asymptotic steady state. In both cases, a good agreement is found again between theory and DSMC results. As a complement of the previous studies, we determine in this paper the dependence of the steady values of the dynamic properties of the suspension on the parameter space of the system. More specifically, we focus on our attention in the temperature ratio $T_1/T_2$ and the fourth degree cumulants $c_1$ and $c_2$ (measuring the departure of the velocity distributions $f_1$ and $f_2$ from their Maxwellian forms). While our approximate theoretical expression for $T_1/T_2$ agree very well with computer simulations, some discrepancies are found for the cumulants. Finally, a linear stability analysis of the steady state solution is also carried out showing that the steady state is always linearly stable.
\end{abstract}

\draft

\date{\today}
\maketitle

\section{Introduction}
\label{sec1}

An effective way of accounting for the influence of the interstitial fluid on the dynamics of solid particles is through a nonconservative external force. \cite{KH01} Usually, for low-Reynolds numbers, this force is composed by two terms: (i) a deterministic drag force proportional to the particle velocity and (ii) a stochastic Langevin-like term. While the first contribution attempts to model background friction (or viscous damping) of grains, the second term mimics the energy gained by the solid particles due to their interactions with the particles of the surrounding molecular gas. The friction of grains on the interstitial gas must not be confused with the static solid body friction which has been shown to play an important role in sheared suspensions. \cite{SMMD13,MS14,SNSPJ20} The suspension model considered here can be also formally derived from the corresponding collision integral by retaining the leading term of the Kramer--Moyal expansion in powers of the mass ratio of the background and solid particles. \cite{RL77,K07,OBB20}

The Navier--Stokes transport coefficients of a binary granular suspension have been recently determined \cite{GKG20} by solving the above suspension model by means of the Chapman--Enskog method \cite{CC70} conveniently adapted to dissipative dynamics. The starting point of this study is the set of Enskog kinetic equations for the mixture with the inclusion of the drag and stochastic forces for each one of the kinetic equations of the components of the mixture. In addition, it is assumed that the state of the surrounding gas is not affected by the presence of the solid particles. It is worthwhile noticing that this suspension model is inspired on simulation results reported in the granular literature \cite{YS09b} where the drift coefficients depend on both the partial and global volume fractions and the mechanical properties of grains (masses and diameters). 

On the other hand, given the intricacies associated with the computation of the transport coefficients in the time-dependent problem, steady-state conditions (namely, when the cooling terms arising from viscous and collisional dissipation are exactly balanced by the heat injected in the system by the bath) were considered to get explicit forms of the diffusion coefficients and the shear and bulk viscosities. The results derived in Ref.\ \onlinecite{GKG20} show that the forms of the diffusion coefficients are in general very different from those found in the case of dry (no gas phase) granular mixtures. \cite{G19} With respect to the shear viscosity, it is found that its form for granular suspensions compare qualitatively well with the one obtained in the dry granular case \cite{G19} for not quite high densities. However, significant quantitative discrepancies between both descriptions (with and without the gas phase) appear for strong inelasticity. The suspension model has been recently \cite{THG21} employed for studying the rheology of a dilute binary mixture of inertial suspension under simple shear flow.

\begin{figure}
	\centering
	\includegraphics[width=1\columnwidth]{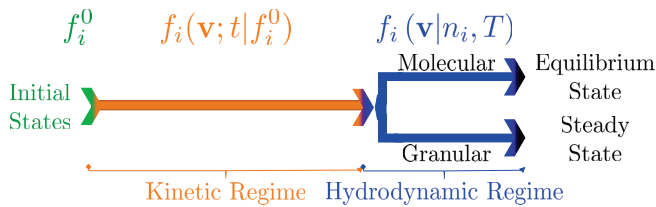}\vspace{-3mm}\\
	\caption{Schematic representation of the time evolution of the distribution functions $f_i$ for molecular and granular mixtures in homogeneous time-dependent states with vanishing mean flow velocity.}
	\label{fig1a}
\end{figure}

A crucial point on the derivation of the Navier--Stokes hydrodynamic equations is the existence of a \emph{normal} (or hydrodynamic) solution \cite{CC70} in the homogeneous problem. This state  is taken in fact as the reference state (zeroth-order approximation) in the Chapman--Enskog expansion around the \emph{local} version of the homogeneous time-dependent state. As widely discussed in different textbooks, \cite{CC70,FK72,GS03} two separate stages can be clearly identified in the evolution of a \emph{molecular} suspension towards equilibrium. First, for times of the order of the mean free time, a \emph{kinetic} stage is identified where the collisions between particles give rise to a relaxation of the distribution function towards a local equilibrium distribution. This kinetic stage depends on the initial preparation of the system. Then, for times much longer than the mean free time, a \emph{hydrodynamic} stage is identified. The hydrodynamic regime is characterized by a slower evolution of the hydrodynamic fields as they approach towards equilibrium. The main feature of the hydrodynamic regime is that the system has practically \emph{forgotten} the details of the initial conditions, except for an implicit dependence on these conditions through the hydrodynamic fields. In the case of granular suspensions, the above two-stage regimes are also expected to be identified, but with the caveat that in the kinetic regime the inelasticity of collisions causes a relaxation towards a non-equilibrium distribution function instead of the local equilibrium distribution. For the sake of clarification, a schematic representation of the two-regime (kinetic and hydrodynamic) evolution of the distribution functions $f_i$ for homogeneous time-dependent states can be found in Fig.\ \ref{fig1a}. 

 Although the applicability of a hydrodynamic description to granular fluids has been supported in the past few years by theory in both the Navier--Stokes \cite{G19} and the non-Newtonian \cite{AS07,As12} regimes, simulations, \cite{BRM01,BRMG02,DHGD02} and experiments, \cite{RBSS02,YHCMW02,HYCMW04} it is interesting to analyze the existence of a hydrodynamic regime in the case of bidisperse granular suspensions. The study of the ``aging to hydrodynamics'' in a multicomponent granular suspension is the first objective of the present work.

We find that, before reaching the stationary regime,  the system ``quickly'' forgets its initial preparation and then evolves towards an unsteady universal (hydrodynamic) state where the velocity distribution function $f_i(\mathbf{v};t)$ of species $i$ has the scaling form
\beq
\label{1.1}
f_i(\mathbf{v};t)=n_i v_0(t)^{-d}\varphi_i\Big(\mathbf{c}(t),T(t)/T_\text{ex}\Big).
\eeq
Here, $n_i$ is the number density of species $i$, $v_0(t)=\sqrt{2T(t)/\overline{m}}$ ($\overline{m}=(m_1+m_2)/2$, $m_i$ being the mass of species $i$) is the thermal speed, $T$ is the global granular temperature, $\mathbf{c}=\mathbf{v}/v_0$, and $T_\text{ex}$ is the (known) background temperature. As in previous studies on driven granular fluids \cite{GMT12,ChVG13} and in contrast to the homogeneous cooling state, \cite{GD99b} the scaling distribution $\varphi_i$ depends on $T$ not only through the dimensionless velocity $\mathbf{c}$ but also on the instantaneous temperature, suitably scaled with respect to the known  bath temperature $T_\text{ex}$. A consequence of the scaling solution \eqref{1.1} is that the velocity moments of $f_i$ evolve in time in a similar form. Thus, for arbitrary initial conditions, one expects that the partial temperatures $T_i(t)/T_\text{ex}$ achieve a universal function (independent of the initial conditions) that depends on time only through the (scaled) temperature $T(t)/T_\text{ex}$. This theoretical result is indeed confirmed here by the direct Monte Carlo simulations (DSMC) \cite{B94} of the Enskog kinetic equation.

The fact that a multicomponent granular suspension admits a hydrodynamic-like type of description opens up possible potential applications. Among them, thermal diffusion segregation of an intruder immersed in a granular suspension is really a very interesting problem. The determination of a segregation criterion will allow us to asses the impact of the interstitial gas on the dynamics of the intruder by comparing this criterion against the one previously reported \cite{G08a,G09,G11} when the gas phase was neglected. 

A surprising and fascinating phenomenon in the transient regime towards the final asymptotic steady state is the so-called Mpemba effect. \cite{MO69} The Mpemba effect is a counterintuitive phenomenon where two samples of fluids at initially different temperatures can evolve in time in such a way that their temperatures cross each other at a given time $t_\text{c}$; the curve for the initially cooler sample stays below the other one for longer times $t>t_\text{c}$. Although this exciting phenomenon was first found in the case of water, similar behaviors to the Mpemba effect have been observed in other systems. \cite{KB20,VD21} However, in spite of the extensive number of works devoted to this problem, the origin of this phenomenon is still unknown. For this reason, different studies based on kinetic theory \cite{LVPS17,TLLVPS19,BPRR20,SP20,THS21,GKG21,GG21,MLLVT21} have been reported in the granular literature for unveiling in a clean way the origin of the Mpemba-like effect (and its inverse counterpart). In the context of molecular suspensions (elastic collisions), we have recently analyzed the Mpemba effect \cite{GKG21,GG21} for initial states close and far away from equilibrium. Theoretical results have been confronted against computer simulations (DSMC and molecular dynamics simulations) showing, in general, an excellent agreement. As a complement of the results reported in Refs.\  \onlinecite{GKG21} and \onlinecite{GG21}, we offer in this paper a quantitative analysis of the Mpemba-like effect for binary granular suspensions, namely, when collisions between solid particles are inelastic. The study of the Mpemba-like effect is the second target of the paper.

As expected, for long times, the suspension reaches an asymptotic stationary state. The study of the dependence of the \emph{steady} values of the dynamic properties of the suspension on the parameter space of the system is the third goal of the present paper. More specifically, we are interested in obtaining the ratio
of kinetic temperatures $T_1/T_2$ and the fourth-degree cumulants $c_1$ and $c_2$ (which measure
non-Gaussian properties of the velocity distributions $f_1$ and $f_2$, respectively) as function of the
mass and diameter ratios, concentration, density, coefficients of restitution, and the background temperature. Theory is compared with DSMC simulations for different systems and coefficients of restitution.  While the theoretical predictions for the temperature ratio compare very well with computer simulations, some discrepancies are found for the cumulants. These discrepancies are of the same order of magnitude as those previously found in dry (no gas phase) granular mixtures. \cite{MG02}

The plan of the paper is as follows. Section \ref{sec2} deals with the Enskog equation of the binary granular suspension for homogeneous time-dependent states. The corresponding evolution equations for the global temperature $T(t)$ and the partial kinetic temperatures $T_i(t)$ (measuring the mean kinetic energy of each species) are also derived from the set of Enskog kinetic equations. Time evolution towards the unsteady hydrodynamic regime is studied in section \ref{sec3} where the existence of the universal hydrodynamic solution \eqref{1.1} is shown at the level of the partial temperatures and the cumulants. Section \ref{sec3} addresses the Mpemba-like effect where exact expressions for the crossing time $t_\text{c}$ and the critical value of the initial temperature differences (which provides information on the occurrence or not of the Mpemba effect) are obtained for initial states close to the asymptotic steady state. A more qualitative analysis is carried out for the so-called large Mpemba-like effect (namely, for initial situations far from the steady state). In both cases (small an large Mpemba effect), theory shows a very good agrement with Monte Carlo simulations. Results for the dynamic properties in the stationary state are studied in section \ref{sec4} while a linear stability analysis of this steady state is also carried out in section \ref{sec5}. The analysis shows that the steady state is always linearly stable. The paper is closed in section \ref{sec6} with a discussion of the results reported here.

\section{Model and kinetic description of binary granular suspensions}
\label{sec2}

Let us consider a granular binary mixture modeled as a gaseous mixture of \emph{inelastic} hard disks ($d=2$) or spheres ($d=3$) of masses $m_1$ and $m_2$ and diameters $\sigma_1$ and $\sigma_2$. For the sake of simplicity, the spheres are assumed to be perfectly smooth and so, collisions among all pairs are characterized by three (positive) constant coefficients of normal restitution $\alpha_{ij}\le 1$ ($i,j=1,2$). The coefficients $\al_{ij}$ can be different for the three types of binary collisions.

\begin{figure}
	\centering
	\includegraphics[width=1\columnwidth]{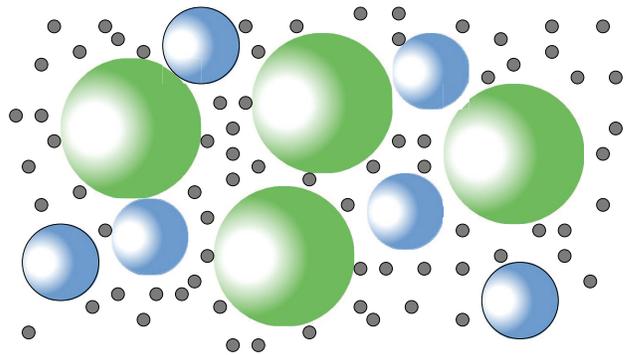}\vspace{-3mm}\\
	\caption{Schematic diagram of the binary suspension. Two kind of particles of masses $m_1$ and $m_2$ are surrounded by a gas of mass $m_g\ll m_{1,2}$. }
	\label{fig1b}
\end{figure}

Grains (solid particles) are immersed in a viscous gas of viscosity $\eta_g\propto \sqrt{T_\text{ex}}$. We assume that the granular mixture is sufficiently rarefied so that, one can suppose that the state of the interstitial fluid (like air or water) is not disturbed by the presence of the solid particles and it can be treated as a \emph{thermostat}. Thus, we assume that both $\eta_g$ and $T_\text{ex}$ are constant quantities. Moreover, as has been widely discussed in previous works, \cite{K90,TK95,SMTK96,WKL03,GFHY16} we also assume that the stresses exerted by the background gas on solid particles are sufficiently weak so they have a small influence on the motion of grains. Thus, the impact of gas phase on collision dynamics can be neglected and consequently, the Enskog--Boltzmann collision operators are not affected by the presence of the interstitial gas. This assumption becomes less reliable as the particle-to-fluid density ratio decreases (for instance, glass beads in liquid water) where one should consider the influence of the gas phase on the collision operator. The use of the kinetic-theory analogy to gas-solid systems is appropriate for relatively massive particles (i.e. high Stokes number) engaging in nearly instantaneous collisions.\cite{GTSH12} These type of systems occur in a wide range of engineering operations, including the riser section of a circulating fluidized bed, pneumatic conveying systems or bubbling fluidized beds. Figure \ref{fig1b} shows a schematic diagram of the system considered in this work. 

Under the above conditions, for moderate densities, the one-particle velocity distribution function $f_i(\mathbf{v},\mathbf{r};t)$ of species or component $i$ of the mixture ($i=1,2$) obeys the set of coupled nonlinear Enskog kinetic equations. For homogeneous and isotropic states, this set reads
\beq
\label{2.1}
\frac{\partial f_i}{\partial t}+\mathcal{F}_i f_i=\sum_{j=1}^2J_{ij}[\mathbf{v}|f_i,f_j],
\eeq
where the Enskog--Boltzmann collision operator $J_{ij}[f_i,f_j]$ is given by
\beqa
\label{2.2}
& &J_{ij}[f_i,f_j]=\sigma_{ij}^{d-1}\chi_{ij}\int \dd\mathbf{v}_2\int \dd\widehat{\boldsymbol{\sigma}}\Theta\left(\widehat{\boldsymbol{\sigma}}\cdot\mathbf{g}_{12}\right)
\left(\widehat{\boldsymbol{\sigma}}\cdot\mathbf{g}_{12}\right)\nonumber\\
& &\times\Big[\alpha_{ij}^{-2}f_i(\mathbf{v}_1'';t)
f_j(\mathbf{v}_2'';t)-f_i(\mathbf{v}_1;t)f_j(\mathbf{v}_2;t)\Big].
\eeqa
Here, $\boldsymbol{\sigma}_{ij}=\sigma_{ij} \widehat{\boldsymbol{\sigma}}$, $\sigma_{ij}=(\sigma_i+\sigma_j)/2$, $\widehat{\boldsymbol{\sigma}}$ is a unit vector directed along the line of centers from the sphere of the component $i$ to that of the component $j$ at contact, $\Theta$ is the Heaviside step function, $\mathbf{g}_{12}=\mathbf{v}_1-\mathbf{v}_2$ is the relative velocity, and $\chi_{ij}(\sigma_{ij})$ is the equilibrium pair correlation function evaluated at contact. The relationship between the pre- and post-collisional velocities is
\beqa
\label{2.3}
\mathbf{v}_1''&=&\mathbf{v}_1-\mu_{ji}\left(1+\alpha_{ij}^{-1}\right)\left(\boldsymbol{\widehat{\sigma}}
\cdot\mathbf{g}_{12}\right)\boldsymbol{\widehat{\sigma}}, \nonumber\\ \mathbf{v}_2'&=&\mathbf{v}_2+\mu_{ij}\left(1+\alpha_{ij}^{-1}\right)\left(\boldsymbol{\widehat{\sigma}}
\cdot\mathbf{g}_{12}\right)\boldsymbol{\widehat{\sigma}},
\eeqa
where $\mu_{ij}=m_i/(m_i+m_j)$.

In Eq.\ \eqref{2.1}, the operator $\mathcal{F}_i$ represents the gas-solid interaction force that models in an effective way the effect of the background viscous gas on the solid particles of the component $i$. For low Reynolds numbers (only laminar flows are considered), this force is usually constituted by two terms: (i) a deterministic viscous drag force proportional to the (instantaneous) particle velocity $\mathbf{v}$ and (ii) a stochastic Langevin-like term that takes into account the effects on a particle of species $i$ coming from neighboring particles. \cite{GTSH12} While the drag force term attempts to account for the loss of energy of particles due to their friction on the surrounding viscous gas (viscous damping), the stochastic term models the energy gained by grains due to their (random) collisions with gas particles. This latter term is represented by a Fokker--Planck collision operator. \cite{WM96} Therefore, the Enskog equation \eqref{2.1} can be written as \cite{GKG20}
\beq
\label{2.4}
\frac{\partial f_i}{\partial t}-\gamma_i\frac{\partial}{\partial\mathbf{v}}\cdot\mathbf{v}f_i-\frac{\gamma_i T_{\text{ex}}}{m_i}\frac{\partial^2 f_i}{\partial v^2}=\sum_{j=1}^2\; J_{ij}[\mathbf{v}|f_i,f_j],
\eeq
where the coefficients $\gamma_i$ are the drag or drift coefficients. Upon writing Eq.\ \eqref{2.4} we have assumed that the mean flow velocity of gas phase vanishes for homogeneous states. Although the drag coefficients $\gamma_i$ should be in general tensorial quantities (as a result of the hydrodynamic interactions between solid particles), here we will assume that those coefficients are scalar quantities independent of configuration of grains. As said in previous works, \cite{HTG17} this simple model is expected to be reliable for describing  inertial suspensions where the mean diameter of suspended particles ranges approximately from 1 to 70 $\mu$m.

On the other hand, lattice-Boltzmann simulations \cite{YS09a,YS09b,HYS10} for binary granular suspensions have shown that the coefficients $\gamma_i$ must be functions of the partial volume fractions
\beq
\label{2.5}
\phi_i=\frac{\pi^{d/2}}{2^{d-1}d\Gamma\left(\frac{d}{2}\right)}n_i\sigma_i^d
\eeq
and the total volume fraction $\phi=\phi_1+\phi_2$. Here, the number density of the component $i$ is defined as
\beq
\label{2.6}
n_i(t)=\int \dd \mathbf{v}\; f_i(\mathbf{v};t).
\eeq
The drag coefficients $\gamma_i$ can be written as $\gamma_i=\gamma_0 R_i$,  where $\gamma_0\propto \eta_g$ and the dimensionless quantities $R_i$ depend on the mole fraction $x_1=n_1/(n_1+n_2)$, the mass ratio $m_1/m_2$, the diameter ratio $\sigma_1/\sigma_2$, and the total volume fraction $\phi$. Although several expressions for the coefficients $\gamma_i$ can be found in the polydisperse gas-solid flows literature, in this work we assume the expression provided in Ref.\ \onlinecite{YS09b} for a three-dimensional system ($d=3$):
\beq
\label{2.7}
\gamma_i=18 \frac{\eta_g}{\rho \sigma_{12}^2} R_i,
\eeq
where $\rho=\rho_1+\rho_2$, $\rho_i=m_in_i$ is the mass density of species $i$, and the dimensionless function $R_i$ is given by
\beqa
\label{2.8}
R_i&=&\frac{\rho \sigma_{12}^2}{\rho_i \sigma_i^2}\frac{(1-\phi)\phi_i\sigma_i}{\phi}\sum_{j=1}^{2}\frac{\phi_j}{\sigma_j}\Bigg[
\frac{10\phi}{\left(1-\phi\right)^2}\nonumber\\
& &+\left(1-\phi\right)^2\left(1+1.5\sqrt{\phi}\right)\Bigg], \quad i=1,2.
\eeqa

In homogeneous states, the properties of primary interest in a binary mixture are the total granular temperature $T(t)$ and the partial temperatures $T_i(t)$ associated with the kinetic energies of each species. They are defined as
\beq
\label{2.9}
T(t)=\sum_{i=1}^2 x_i T_i(t), \quad T_i(t)=\frac{1}{dn_i(t)}\int \dd \mathbf{v}\; m_i v^2\; f_i(\mathbf{v};t),
\eeq
where $x_2=1-x_1$. The time dependence of $T(t)$ and $T_i(t)$ follows from the set of Enskog equations \eqref{2.1} that gives \cite{GKG20}
\beq
\label{2.10}
\frac{\partial T}{\partial t}=2\sum_{i=1}^2 x_i\gamma_i\left(T_\text{ex}-T_i\right)-\zeta T,
\eeq
\beq
\label{2.10.1}
\frac{\partial T_i}{\partial t}=2\gamma_i\left(T_\text{ex}-T_i\right)-\zeta_i T_i,
\eeq
where $\zeta_i$ is the cooling rate associated with $T_i$ and $\zeta$ is the total cooling rate. The latter quantity gives the rate of change of the total kinetic energy due to inelastic collisions among all components of the mixture. The cooling rates $\zeta$ and $\zeta_i$ are defined, respectively, as
\beq
\label{2.11}
\zeta=\frac{1}{T}\sum_{i=1}^2 x_i T_i\zeta_i, \quad  \zeta_i=-\frac{m_i}{dn_iT_i}\sum_{j=1}^2\int\dd\mathbf{v}\; v^2 J_{ij}[\mathbf{v}|f_i,f_j].
\eeq

Equation  \eqref{2.10} shows the competing mechanisms appearing in the evolution of the granular temperature towards its steady state $T_\text{s}=\lim_{t\to \infty}T(t)$. Thus, the stationary temperature is approached from below ($T(t)<T_\text{s}$) when the heat supplied by the external bath ($2\sum_i x_i \gamma_i T_\text{ex}$) prevails over the cooling terms arising from viscous friction ($2\sum_i x_i \gamma_i T_i$) and collisional cooling ($\zeta T$); this situation will be referred to as the \emph{heating} case. Otherwise, the stationary temperature is achieved from above ($T(t)>T_\text{s}$) and this will be referred to as the \emph{cooling} case. The interesting question is if an \emph{unsteady} hydrodynamic regime exists in both situations (heating and/or cooling cases) before the granular binary suspension achieves the asymptotic steady state.

\section{Time evolution towards the stationary state: the unsteady hydrodynamic regime}
\label{sec3}

In order to analyze the homogeneous transient regime throughout the evolution of $T(t)$ and $T_i(t)$, it is convenient to introduce dimensionless variables for temperature and time. Let us define the reduced temperatures $\theta(t)=T(t)/T_\text{ex}$ and $\theta_i(t)=T_i(t)/T_\text{ex}$, the reduced friction coefficients $\gamma_i^*(t)=\gamma_i/\nu(t)$ and the reduced cooling rates $\zeta^*(t)=\zeta(t)/\nu(t)$ and $\zeta_i^*(t)=\zeta_i(t)/\nu(t)$. Here, the effective collision frequency $\nu(t)$ is defined as
\beq
\label{3.1}
\nu(t)=n\sigma_{12}^{d-1}v_0(t),
\eeq
where $n=n_1+n_2$ is the total number density of the mixture and we recall that $v_0(t)=\sqrt{2T(t)/\overline{m}}$. According to Eqs.\ \eqref{2.7} and \eqref{2.8}, the dimensionless drag coefficients $\gamma_i^*$ can be expressed more explicitly in terms of the dimensionless functions $R_i$ and the (reduced) temperature $\theta$ as
\beq
\label{3.2}
\gamma_i^*=\lambda_i\theta^{-1/2},\qquad \lambda_i=\frac{\sqrt{2}\pi^{d/2}}{2^dd\Gamma\left(\frac{d}{2}\right)}\frac{R_i}{\sqrt{T_\text{ex}^*}\sum_j(\sigma_{12}/\sigma_j)^d
	\phi_j},
\eeq
where
\beq
\label{3.3}
T_\text{ex}^*\equiv \frac{T_\text{ex}}{\overline{m}\sigma_{12}^2\gamma_0^2}
\eeq
is the reduced background temperature. In terms of the above dimensionless quantities, Eqs.\ \eqref{2.10} and \eqref{2.10.1} can be written as
\beq
\label{3.4}
\frac{\partial \theta}{\partial t^*}=2\sum_{i=1}^2x_i\lambda_i(1-\theta_i)-\theta^{3/2}\zeta^*,
\eeq
\beq
\label{3.5}
\frac{\partial \theta_i}{\partial t^*}=2\lambda_i(1-\theta_i)-\theta^{1/2}\zeta_i^*\theta_i,
\eeq
where the reduced time $t^*=n\sigma_{12}^{d-1}\sqrt{2T_\text{ex}/\overline{m}}t$ and $\zeta^*=\theta^{-1}\left(x_1 \theta_1 \zeta_1^*+x_2 \theta_2 \zeta_2^*\right)$.

It is quite apparent that to solve the Enskog kinetic equations \eqref{2.4} one has to provide specific initial conditions $f_i(\mathbf{v};0)\equiv f_i^{0}(\mathbf{v})$. In this sense, the solution $f_i(\mathbf{v};t)$ can be considered as a functional of the initial distribution, namely, $f_i(\mathbf{v};t)=f_i(\mathbf{v};t|f_i^{0})$. \cite{AS07} Analogously, the velocity moments of $f_i$ (such as the partial temperatures $\theta_i$) are also functionals of the initial distribution. Since the only time-dependent hydrodynamic variable in the homogeneous state is the granular temperature, for times longer than the mean free time, the existence of the hydrodynamic regime necessarily implies that the time-dependence of the distribution function $f_i(\mathbf{v};t)$ is through the temperature $T(t)$. It follows from dimensional analysis that $f_i(\mathbf{v};t)$ has the scaling form \eqref{1.1}, i.e.,
\beq
\label{3.6}
f_i(\mathbf{v};t|f_i^{0})\to n_i v_0(t)^{-d}\varphi_{i}(\mathbf{c}(t),\theta(t)),
\eeq
where we recall that $\mathbf{c}(t)\equiv \mathbf{v}/v_0(t)$ is the particle velocity expressed in units of the time-dependent thermal speed. Upon writing the right hand side of Eq.\ \eqref{3.6} we have accounted for that $\gamma_i^*(t)$ depends on time only through its dependence on $\theta(t)$. For given values of the parameters of the mixture (concentration, masses, sizes, density, and coefficients of restitution), the scaled distribution $\varphi_{i}(\mathbf{c}(t),\theta(t))$ is a universal function independent of the initial distribution $f_i^{0}$; its time-dependence is enclosed not only in the dimensionless velocity $\mathbf{c}$ but also in the scaled temperature $\theta$. The fact that the velocity statistics is envisioned by a two-parameter scaling form (at a variance with the homogenous cooling state in undriven granular mixtures \cite{GD99b,G19}) is a common feature in driven granular gases. \cite{GMT12,ChVG13} Thus, if an unsteady hydrodynamic description exists, the different solutions $f_i(\mathbf{v};t|f_i^{0})$ to the set of Enskog equations \eqref{2.4} must \emph{collapse} in the universal form \eqref{3.6}. Then, for very long times, the steady state is eventually achieved where $\varphi_{i}(\mathbf{c},\theta)\to \varphi_{i}(\mathbf{c},\theta_\text{s})$, $\theta_\text{s}$ being the stationary value of the (reduced) temperature. A consequence of Eq.\ \eqref{3.6} is that the velocity moments of the distribution $f_i(\mathbf{v};t)$ will evolve in a similar way. In particular, regardless of the initial state, the partial temperature $\theta_i(t|f_i^{0})$ will be \emph{attracted} by the universal function $\theta_i(\theta(t))$.

\begin{figure}
	\centering
	\includegraphics[width=0.9\columnwidth]{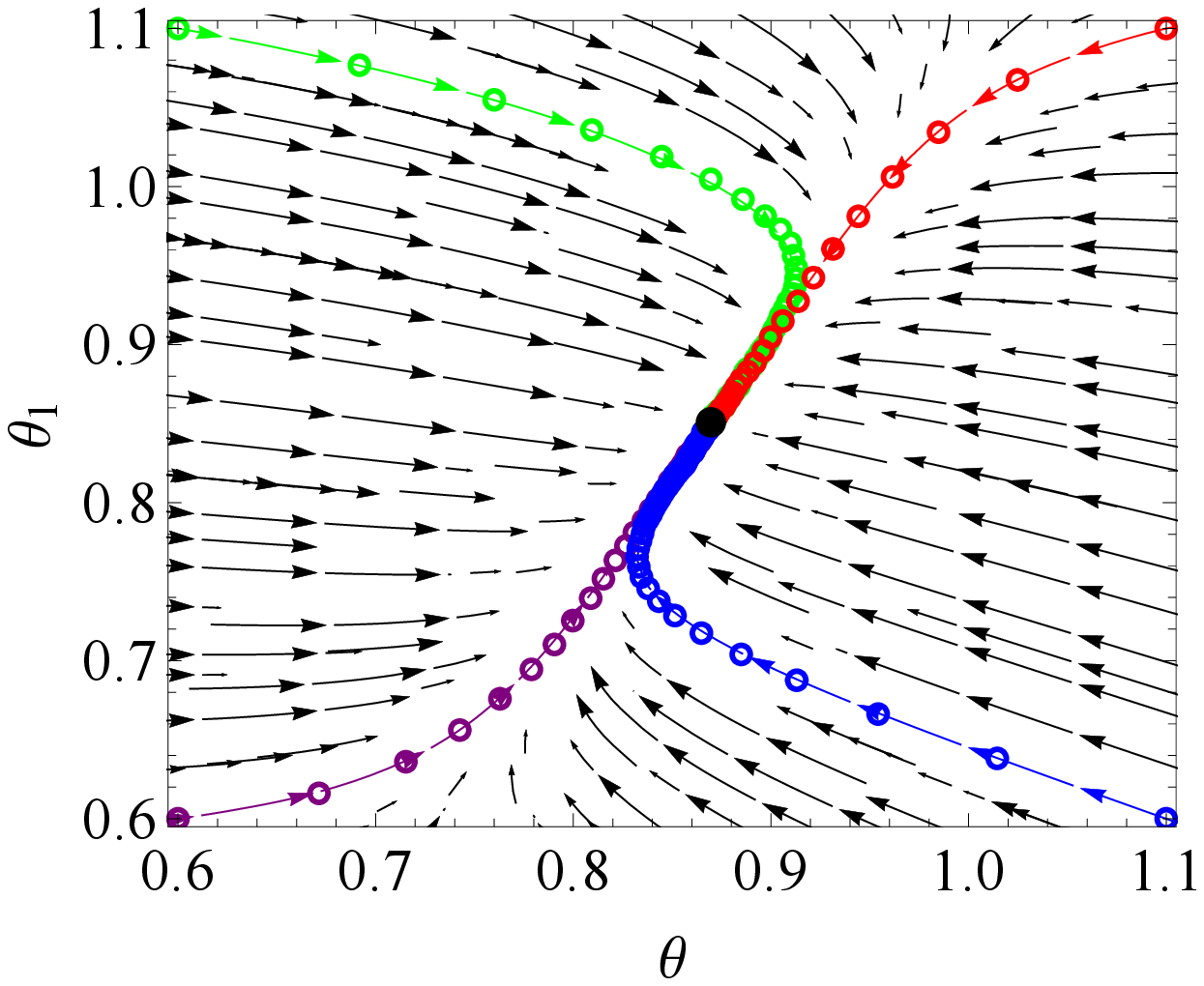}\vspace{5mm}\\
	\includegraphics[width=0.9\columnwidth]{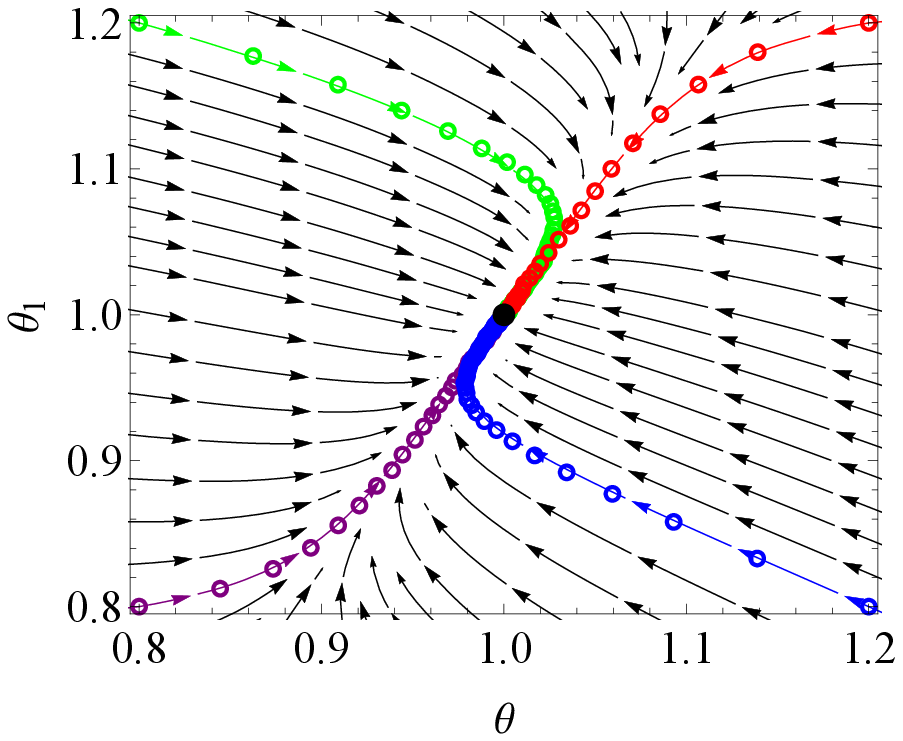}
	\caption{Evolution of the (reduced) partial temperature $\theta_1(t^*)$ versus the (reduced) temperature $\theta(t^*)$ for  $m_1/m_2=10$, $\sigma_1/\sigma_2=1$, $x_1=\frac{1}{2}$, and a common coefficient of restitution $\alpha$ ($\alpha\equiv\alpha_{11}=\alpha_{12}=\alpha_{22}$). Solid lines represent the theoretical values and symbols DSMC data. Top panel corresponds to $\alpha=0.9$ and bottom panel to $\alpha=1$. Top panel: the initial values $\theta_1(\theta)$ of the colored lines are $\theta_1(0.6)=0.6$ (purple line and symbols), $\theta_1(0.6)=1.1$ (green line and symbols), $\theta_1(1.1)=0.6$ (blue line and symbols), and $\theta_1(1.1)=1.1$ (red line and symbols). Bottom panel: the initial values $\theta_1(\theta)$ of the colored lines are $\theta_1(0.8)=0.8$ (purple line and symbols), $\theta_1(0.8)=1.2$ (green line and symbols), $\theta_1(1.2)=0.8$ (blue line and symbols), and $\theta_1(1.2)=1.2$ (red line and symbols).  The remaining parameters are $d=3$, $\phi=0.1$, and  $T_\text{ex}^*=1$. The filled circles correspond to the values of $\theta_1$ in the steady state.}
	\label{fig1}
\end{figure}

On the other hand, according to Eqs.\ \eqref{3.4} and \eqref{3.5}, to confirm the existence of the hydrodynamic solution one needs to know the partial cooling rates $\zeta_i^*$, which are defined by Eq.\ \eqref{2.11} in terms of the velocity distributions $f_i(\mathbf{v};t)$. Here, to estimate $\zeta_i^*$ we take the simplest approximation for the distributions $f_{i}(\mathbf{v};t)$, namely, the Maxwellian distributions $f_{i,\text{M}}(\mathbf{v};t)$ defined with the partial temperatures $T_i(t)$:
\beq
\label{3.7}
f_{i,\text{M}} (\mathbf{v};t)=n_i \left(\frac{m_i}{2\pi k_B T_i(t)}\right)^{d/2} \exp\left(-\frac{m_i v^2}{2k_BT_i(t)}\right).
\eeq
In this approximation, the (reduced) partial cooling rates $\zeta_i^*$ are given by \cite{G19}
\beqa
\label{3.8}
\zeta_i^*&=&\frac{4\pi^{(d-1)/2}}{d\Gamma\left(\frac{d}{2}\right)}\sum_{j=1}^2 x_j \chi_{ij}\mu_{ji}\left(\frac{\sigma_{ij}}{\sigma_{12}}\right)^{d-1}\left(\frac{\beta_i+\beta_j}{\beta_i\beta_j}\right)^{1/2}
\nonumber\\
& &\times(1+\alpha_{ij})\left[1-\frac{\mu_{ji}}{2}(1+\alpha_{ij})\frac{\beta_i+\beta_j}{\beta_j}\right],	
\eeqa
where $\beta_i=M_i\theta/\theta_i$ and $M_i=m_i/\overline{m}$. In addition, to make a plot $\theta_i(t)$ versus $\theta(t)$, the form of the pair correlation function is also needed. A good approximation for $\chi_{ij}$ for spheres ($d=3$) is \cite{GH72,LL73}
\beq
\label{3.9}
\chi_{ij}=\frac{1}{1-\phi}+\frac{3}{2}\frac{\phi}{(1-\phi)^2}\frac{\sigma_i\sigma_jM_2}{\sigma_{ij}M_3}+\frac{1}{2}
\frac{\phi^2}{(1-\phi)^3}\left(\frac{\sigma_i\sigma_jM_2}{\sigma_{ij}M_3}\right)^2,
\eeq
where $M_\ell=\sum_i x_i\sigma_i^\ell$. A parametric plot $\theta_1(t^*)$ versus $\theta(t^*)$ is a quite useful test to see if actually an unsteady hydrodynamic regime is established, namely, if $\theta_1(t^*)\to \theta_1(\theta(t^*))$, where the function $\theta_1(\theta)$ must be independent of the initial conditions. Such a parametric plot is shown in Fig.\ \ref{fig1} for a binary mixture with parameters $\sigma_1/\sigma_2=1$, $m_1/m_2=10$, $x_1=\frac{1}{2}$, $\phi=0.1$, and $T_\text{ex}^*=1$. Two different values of the (common) coefficient of restitution $\al_{ij}\equiv \al$ are considered: $\al=1$ (elastic collisions) and $\al=0.9$ (inelastic collisions). Different cooling [$\theta(t^*)$ decreases in time] and heating [$\theta(t^*)$ increases in time] cases have been considered in Fig.\ \ref{fig1}. Lines are the theoretical results derived by numerically solving Eqs.\ \eqref{3.4} and \eqref{3.5} with the Maxwellian approximation \eqref{3.8} for $\zeta_i^*$ while symbols refer to the results obtained via DSMC simulations. Figure \ref{fig1} highlights that, for sufficiently long times, the different curves (corresponding to different initial conditions) are attracted to a common universal curve (time-dependent hydrodynamic regime) where $\theta_1$ depends on time through the granular temperature $\theta$ only. Moreover, an excellent agreement is found between the theoretical and the DSMC results in both granular and elastic cases. Although not shown here, similar results are found for smaller values of $\alpha$ ($\alpha \lesssim 0.5$).

\subsection{Unsteady hydrodynamic regime. Leading Sonine approximation}

In the unsteady hydrodynamic regime, $f_i$ adopts the hydrodynamic form \eqref{3.6} and so, the Enskog equation \eqref{2.4} for the scaled distributions $\varphi_i(\mathbf{c}, \theta)$ reads
\beqa
\label{3.10}
& &\left[2\sum_{i=1}^2x_i\gamma_i^*(1-\theta_i)-\zeta^*\theta\right]\frac{\partial \varphi_i}{\partial \theta}\nonumber\\& &+\Bigg[\frac{\zeta^*}{2}-\sum_{i=1}^2x_i\gamma_i^*\theta^{-1}
(1-\theta_i)-\gamma_i^*\Bigg]\frac{\partial}{\partial\mathbf{c}}\cdot\mathbf{c}\varphi_i\nonumber\\& &-\frac{\gamma^*_i}{2M_i\theta}\frac{\partial^2\varphi_i}{\partial c^2}=\sum_{j=1}^2\; J^*_{ij}[\mathbf{c}|\varphi_i,\varphi_j],
\eeqa
where $J^*_{ij}= \ell J_{ij}/(n_i v_0^{1-d})$ and use has been made of the property \cite{GKG20}
\beq
\label{3.11}
T\frac{\partial f_i}{\partial T}=-\frac{1}{2}\frac{\partial}{\partial \mathbf{v}}\cdot \mathbf{v}f_i+n_iv_0^{-d}\theta\frac{\partial\varphi_i}{\partial\theta}.
\eeq
Here, the derivative $\partial\varphi_i/\partial\theta$ is taken at constant $\mathbf{c}$. In addition, in the hydrodynamic regime, the evolution equation \eqref{3.5} can be rewritten as
\beq
\label{3.12}
\Lambda\frac{\partial \theta_i}{\partial \theta}=\Lambda_i, \quad \Lambda_i=2\gamma_i^*\left(1-\theta_i\right)-\theta_i\zeta_i^*, \quad \Lambda=x_1\Lambda_1+x_2\Lambda_2.
\eeq

The exact solution to the time-dependent Enskog equation \eqref{3.10} is not known to date. Although we have seen before that the Maxwellian distribution \eqref{3.7} yields a good estimate of the partial cooling rates $\zeta_i^*$, the scaled distribution $\varphi_i(\mathbf{c})$ differs from its Maxwellian form
\beq
\label{3.13}
\varphi_{i,\text{M}}(\mathbf{c})=\pi^{-d/2}\beta_i^{d/2}\operatorname{e}^{-\beta_ic^2}.
\eeq
An usual way of assessing the deviations of $\varphi_i$ from $\varphi_{i,\text{M}}$ in the range of low and intermediate velocities is to expand $\varphi_i$ in a complete set of Laguerre (or Sonine) polynomials where the coefficients (or cumulants) $c_i$ of such an expansion are the velocity moments of the distribution. Based on the assumption that the cumulants $c_i$ are small, approximate expressions for them can be achieved by truncating the series expansion at a given order.  Hence, the leading Sonine approximation to $\varphi_i$ is given by
\beq
\label{3.13b}
\varphi_i(\mathbf{c})=\varphi_{i,\text{M}}(\mathbf{c})\left\{1+\frac{c_i}{2}\left[\beta_i^2c^4-(d+2)\beta_ic^2+\frac{d(d+2)}{4}\right]\right\},
\eeq
where the fourth-degree cumulants $c_i$ are defined as
\beq
\label{3.14}
c_i=\frac{4}{d(d+2)}\beta_i^2\int\dd\mathbf{c}\; c^4\varphi_i(\mathbf{c})-1.
\eeq

\begin{figure}
	\centering
	\includegraphics[width=0.9\columnwidth]{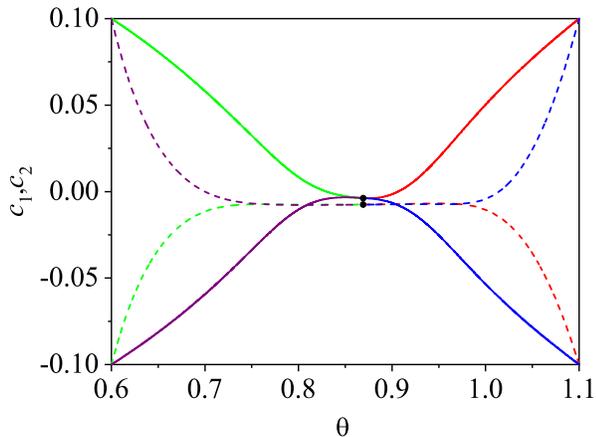}
	\caption{Evolution of the cumulants $c_1$ and $c_2$ versus the (reduced) temperature $\theta$ for  $m_1/m_2=10$, $\sigma_1/\sigma_2=1$, $x_1=\frac{1}{2}$, and a common coefficient of restitution $\alpha=0.9$ ($\alpha\equiv\alpha_{11}=\alpha_{12}=\alpha_{22}$). Solid and dashed lines represent the time evolution of $c_1$ and $c_2$, respectively. The initial values $\left\{\theta, \theta_1, c_1, c_2\right\}$ of the colored lines are $\left\{0.6,0.6,0.1,-0.1\right\}$ (green line and symbols), $\left\{0.6,0.6,-0.1,0.1\right\}$ (purple line and symbols), $\left\{1.1,1.1,0.1,-0.1\right\}$ (red line and symbols), and $\left\{0.6,0.6,-0.1,0.1\right\}$ (blue line and symbols). The remaining parameters are $d=3$, $\phi=0.1$, and  $T_\text{ex}^*=1$. The filled circles correspond to the values of $c_1$ and $c_2$ in the steady state.}
	\label{fig2}
\end{figure}

We want to analyze the time dependence of the coefficients $c_i$, or equivalently, the dependence of $c_i$ on $\theta$. To obtain self-consistent results, the partial cooling rates $\zeta_i^*$ are now estimated by using the leading Sonine polynomial term \eqref{3.13b}. Thus, according to the constraint $\theta(t^*)=x_1\theta_1(t^*)+x_2\theta_2(t^*)$, the unknown (independent) quantities are the partial temperature $\theta_1$ and the cumulants $c_1$ and $c_2$. The equation governing the time evolution of $\theta_1$ is given by Eq.\ \eqref{3.12} with $i=1$. The time evolution equations for the cumulants can be obtained by multiplying the set of Enskog equations \eqref{3.10} by $c^4$ and integrating over $\mathbf{c}$. After some algebra, one gets
\beq
\label{3.15}
\Lambda\frac{\partial c_i}{\partial\theta}+2\left(\Lambda_i\theta_i^{-1}+2\gamma_i^*\right)\left(1+c_i\right)
-4\gamma_i^*\theta_i^{-1}=\frac{4\beta_i^2}{d(d+2)}\Sigma_i,
\eeq
where
\beq
\label{3.16}
\Sigma_i=\sum_{j=1}^2\int\dd\mathbf{v}\; c^4 J_{ij}^*[\mathbf{c}|\varphi_i,\varphi_j].
\eeq
The partial cooling rates $\zeta_i^*$ as well as the collisional moments $\Sigma_i$ are obtained by substituting the leading Sonine approximation \eqref{3.13b} into Eqs.\ \eqref{2.11} and \eqref{3.16}, retaining only linear terms in $c_i$, and integrating over velocity. The final expressions can be written as \cite{KG14}
\beq
\label{3.17}
\zeta_1^*=\zeta_{10}+\zeta_{11}c_1+\zeta_{12}c_2, \quad \zeta_2^*=\zeta_{20}+\zeta_{22}c_2+\zeta_{21}c_1,
\eeq
\beq\label{3.18}
\Sigma_1=\Sigma_{10}+\Sigma_{11}c_1+\Sigma_{12}c_2, \quad \Sigma_1=\Sigma_{20}+\Sigma_{22}c_2+\Sigma_{21}c_1,
\eeq
where the explicit forms of $\zeta_{ij}$ and $\Sigma_{ij}$ are displayed in the Appendix for the sake of completeness.

Figure \ref{fig2} illustrates the dependence of the coefficients $c_1$ and $c_2$ on $\theta$ for the same initial conditions as in Fig.\ \ref{fig1}. It is quite apparent that, after a transient period, the cumulants converge towards the universal hydrodynamic regime in the same way as the partial temperatures $\theta_i$ do. We also observe that the temporal duration of the unsteady hydrodynamic regime of the cumulant $c_2$ is greater than that of the cumulant $c_1$.

\section{Mpemba-like effect in binary granular suspensions}
\label{sec4}

As mentioned in section \ref{sec1}, before considering steady situations, it is interesting to analyze the so-called Mpemba-like effect in binary granular suspensions. Mpemba-like effect is a counterintuitive phenomenon where an initially hotter sample can cool down sooner than the colder one. This effect was experimentally observed for the first time many years ago by E. B. Mpemba \cite{MO69} in the case of water. Although different mechanisms have been proposed in the literature to explain the Mpemba effect in such a system, \cite{ERS08,K09bis,JG15,IC16,KPZSN19,KRHV19,GLH19} the problem is still open since there are still doubts about the origin of this exciting phenomenon. \cite{BL16,BL20} For this reason, to gain some insight into this complex problem, kinetic theory tools have been widely employed in the last few years to understand the cause of a reduction in the relaxation time as the trigger of the Mpemba-like effect in molecular \cite{SP20,GKG21} and granular \cite{LVPS17,TLLVPS19,BPRR20,THS21,MLLVT21} gases. In particular, we have recently analyzed \cite{GKG21,GG21} this phenomenon (and its inverse and mixed counterparts) in the case of molecular binary mixtures driven by a stochastic bath with friction. Theoretical approximate results have been confronted against computer simulations showing an excellent agreement. Although some preliminary results for inelastic collisions were also reported Ref.\ \onlinecite{GKG21}, we complement in this section the results obtained before by offering a more quantitative analysis of the Mpemba-like effect in binary granular suspensions.

Let us assume two identical homogeneous states A and B except for their initial values of the reduced global temperatures $\theta_\text{A}^{(0)}$ and $\theta_\text{B}^{(0)}$ and their reduced partial temperatures $\theta_{1,\text{A}}^{(0)}$ and $\theta_{1,\text{B}}^{(0)}$ \footnote{Here, in contrast to the analysis made in Ref.\ \onlinecite{GKG21}, the temperature $\theta_1$ is employed instead of the temperature ratio $\theta_1/\theta_2$ for studying the Mpemba-like effect. The results are, of course, equivalent in both descriptions.}. As discussed in Ref.\ \onlinecite{GKG21}, the fact that the time evolution equations obeying $\theta(t^*)$ and $\theta_1(t^*)$ are coupled [see Eqs.\ \eqref{3.4} and \eqref{3.5}] opens up the possibility that $\theta_\text{A}(t^*_\text{c})=\theta_\text{B}(t^*_\text{c})$ at a given crossing time $t^*_\text{c}$ (Mpemba-like effect) before reaching the (common) asymptotic steady state value $\theta_\text{s}$.

To analyze the time evolution of $\theta$ and $\theta_1$, let us rewrite Eqs.\ \eqref{3.4} and \eqref{3.5} as
\beq
\label{4.1}
\frac{\partial \theta}{\partial t^*}=\Phi(\theta,\theta_1), \quad  \frac{\partial \theta_1}{\partial t^*}=\Psi(\theta,\theta_1)
\eeq
where
\beqa
\label{4.2}
\Phi(\theta,\theta_1)&=&\Phi_1+\Phi_2(\theta)+\Phi_3(\theta_1)+\Phi_4(\theta,\theta_1), \nonumber\\ \Psi(\theta,\theta_1)&=&\Psi_1+\Psi_2(\theta)+\Psi_3(\theta,\theta_1).
\eeqa
Here, we have introduced the following quantities
\begin{align}
\label{4.3}
\Phi_1&=2\left(x_1\lambda_1+x_2\lambda_2\right), \quad \Phi_2(\theta)=-2\lambda_2 \theta,\nonumber\\
\Phi_3(\theta_1)&=-2x_1\left(\lambda_1-\lambda_2\right)\theta_1, \nonumber\\ \Phi_4(\theta,\theta_1)&=-\theta^{1/2}\left[x_1\theta_1\left(\zeta_1^*-\zeta_2^*\right)+\theta \zeta_2^*\right],
\end{align}

\beq
\label{4.4}
\Psi_1=2\lambda_1, \quad \Psi_2(\theta_1)=-2\lambda_1 \theta_1, \quad \Psi_3(\theta,\theta_1)=-\theta^{1/2}\theta_1 \zeta_1^*.
\eeq

In contrast to other memory effects reported in the case of molecular and granular gases, \cite{LVPS17,SP20,MLLVT21} here we use the partial temperature as the kinetic variable whose evolution couples with that of the temperature. For this reason, no cumulants are needed in the description of the Mpemba-like effect. Thus, in order to solve Eqs.\ \eqref{4.3} and \eqref{4.4}, since the impact of the cumulants $c_i$ on the partial temperatures $\theta_i$ is very small, we will neglect them for the sake of simplicity to estimate the partial cooling rates $\zeta_i^*$. In that case, according to Eq. \ \eqref{3.8}, $\zeta_i^*$ can be rewritten as
\beq
\label{4.5}
\zeta_1^*=\sqrt{\frac{\theta_1}{M_1\theta}}\zeta_1'(\beta),
\eeq
where
\beqa
\label{4.6}
\zeta_1'(\beta)&=&\frac{\sqrt{2}\pi^{(d-1)/2}}{d\Gamma\left(\frac{d}{2}\right)}x_1\chi_{11}\left(\frac{\sigma_1}{\sigma_{12}}\right)^{d-1}
(1-\alpha_{11}^2)\nonumber\\& &+\frac{4\pi^{(d-1)/2}}{d\Gamma\left(\frac{d}{2}\right)}x_2\chi_{12}\mu_{21}(1+\beta)^{1/2}(1+\alpha_{12})\nonumber\\
& &\times\left[1-\frac{\mu_{21}}{2}(1+\alpha_{12})(1+\beta)\right].
\eeqa
Here, $\beta=\beta_1/\beta_2=m_1\theta_2/m_2\theta_1$. The expression of $\zeta_2^*$ can be easily obtained from Eqs.\ \eqref{4.5} and \eqref{4.6} by interchanging 1 and 2 and setting $\beta\rightarrow \beta^{-1}$.

For elastic collisions ($\al_{ij}=1$), $\theta\zeta^*=x_1\theta_1\left(\zeta_1^*-\zeta_2^*\right)+\theta \zeta_2^*=0$ and so, $\Phi_4=0$ according to the last identity in Eq.\ \eqref{4.3}. Thus, for molecular mixtures, the study of the Mpemba effect becomes more simple since the time evolution of $\theta(t^*)$ is essentially ruled by the function $\Phi_2(\theta)+\Phi_3(\theta_1)$. On the other hand, for inelastic collisions ($\Phi_4 \neq 0$), the analysis of the Mpemba effect is much more intricate than for molecular mixtures. Thus, in order to offer a quantitative analysis, we consider first initial states which are very close to the final steady state. This will allow us to get explicit expression for the crossing time $t_\text{c}^*$ and, as a consequence, for the initial conditions needed for the crossover in the evolution of the temperatures of the two samples. In this context, the set of coupled differential equations \eqref{4.2} can be linearized around the stationary solutions $\theta_\text{s}$ and $\theta_{1,\text{s}}$, where the subscript $s$ means that the quantity is evaluated in the steady state. An exhaustive study of the dependence of $\theta_\text{s}$ and $\theta_{1,\text{s}}$ on the parameter space will be provided in section \ref{sec5}.


We want to solve the set of equations \eqref{4.1}--\eqref{4.2} by assuming small deviations from the steady state solution. Therefore, we write
\beq
\label{4.7}
\theta(t^*)=\theta_\text{s}+\delta \theta(t^*), \quad \theta_1(t^*)=\theta_{1,\text{s}}+\delta \theta_1(t^*).
\eeq
Substitution of Eqs.\ \eqref{4.7} into Eqs.\ \eqref{4.2} and retaining only linear terms in $\delta\theta$ and $\delta\theta_1$, one obtains the set of linear differential equations
\beq
\label{4.8}
\frac{\partial}{\partial t^*}
\begin{pmatrix} \delta \theta \\ \delta \theta_1 \end{pmatrix}=\mathcal{L}\cdot \begin{pmatrix} \delta \theta \\ \delta \theta_1 \end{pmatrix}.
\eeq
The square matrix $\mathcal{L}$ is composed by the following elements:
\begin{widetext}
\beq
\label{4.9}
\mathcal{L}_{11}=-2\lambda_2-\frac{3}{2}\theta_{\text{s}}^{1/2}\zeta_{2,\text{s}}^*-\theta_{1,\text{s}}^{1/2}
\frac{x_1M_1^{1/2}}{x_2M_2}\left(\frac{\partial\zeta_1'}{\partial\beta}\right)_\text{s}+\frac{\theta_{2,\text{s}}^{3/2}}
{\theta_{1,\text{s}}}\frac{M_1}{M_2^{3/2}}\left(\frac{\partial\zeta'_2}{\partial\beta}\right)_\text{s},
\eeq
\beq
\label{4.10}
\mathcal{L}_{12}=-2x_1\left(\lambda_1-\lambda_2\right)+\frac{3}{2}\theta_{\text{s}}^{1/2}x_1\left(\zeta_{2,\text{s}}^*-\zeta_{1,\text{s}}^*\right)
+\frac{\theta_{\text{s}}}{\theta_{1,\text{s}}^{1/2}}\frac{x_1 M_1^{1/2}}{x_2M_2}
\left(\frac{\partial\zeta_1'}{\partial\beta}\right)_\text{s}+\frac{\theta_\text{s}\theta_{2,\text{s}}^{3/2}}{\theta_{1,\text{s}}^2}
\frac{M_1}{M_2^{3/2}}\left(\frac{\partial\zeta'_2}{\partial\beta}\right)_\text{s},
\eeq
\beq
\label{4.11}
\mathcal{L}_{21}=-\theta_{1,\text{s}}^{1/2}
\frac{M_1^{1/2}}{x_2M_2}\left(\frac{\partial\zeta'_1}{\partial\beta}\right)_\text{s}, \quad
\mathcal{L}_{22}=-2\lambda_1-\frac{3}{2}\theta_{\text{s}}^{1/2}\zeta_{1,\text{s}}^*+\theta_{\text{s}}\theta_{1,\text{s}}^{-1/2}
\frac{M_1^{1/2}}{x_2M_2}\left(\frac{\partial\zeta'_1}{\partial\beta}\right)_\text{s}.
\eeq
Here, the derivatives of $\zeta_1'$ and $\zeta_2'$ on $\beta$ are evaluated in the steady state.
\end{widetext}
The solution of the matrix equation \eqref{4.8} for $\delta \theta(t^*)$ is
\beqa
\label{4.12}
\delta \theta(t^*)&=&\frac{1}{\lambda_{+}-\lambda_{-}}\bigg\{\left[\left(\mathcal L_{11}-\lambda_{-}\right)\delta \theta_0+\mathcal L_{12}\delta\theta_{1,0}\right]\operatorname{e}^{\lambda_{+}t^*}\nonumber\\
& &+\left[\left(\lambda_{+}-\mathcal L_{11}\right)\delta \theta_0-\mathcal L_{12}\delta\theta_{1,0}\right]\operatorname{e}^{\lambda_{-}t^*}\bigg\},
\eeqa
where $\delta \theta_0$ and $\delta \theta_{1,0}$ are the initial values of $\delta \theta$ and $\delta \theta_1$, respectively. The eigenvalues of the matrix $\mathcal{L}$ are given by
\beq
\label{4.13}
\lambda_{\pm}=\frac{1}{2}\left[\mathcal L_{11}+\mathcal L_{22}\pm\sqrt{\left(\mathcal L_{11}-\mathcal L_{22}\right)^2+4\mathcal L_{12}\mathcal L_{21}}\right].
\eeq

Let us assume that the initial temperature of the state A is larger than that of the state B ($\theta_\text{A}^{(0)}>\theta_\text{B}^{(0)}$). The possible crossing time $t_\text{c}^*$ for the occurrence of the Mpemba effect can be obtained from the condition $\delta \theta_\text{A}(t_c^*)=\delta \theta_\text{B}(t_\text{c}^*)$. This leads to the result
\beq
\label{4.14}
t^*_\text{c}=\frac{1}{\lambda_{-}-\lambda_{+}}\ln \frac{\mathcal L_{12}+(\mathcal L_{11}-\lambda_{-})\Delta \theta_0/\Delta \theta_{1,0}}{\mathcal L_{12}-(\lambda_+-\mathcal L_{11})\Delta \theta_0/\Delta \theta_{1,0}},
\eeq
where $\Delta \theta_0=\theta_\text{A}^{(0)}-\theta_\text{B}^{(0)}$ and $\Delta\theta_{1,0}=\theta_{1,\text{A}}^{(0)}-\theta_{1,\text{B}}^{(0)}$. As expected, \cite{GKG21} in the linear theory, for given values of the parameters of the mixture, $t^*_\text{c}$ depends on the initial conditions \emph{only} through the single control parameter $\Delta \theta_0/\Delta \theta_{1,0}$. Moreover, since $\lambda_--\lambda_+<0$ and $t^*_\text{c}\in \mathbb{R}^+$, the argument of the logarithm in Eq.\ \eqref{4.14} belongs to the interval $(0,1)$. According to this constraint, the initial values must satisfy the conditions \cite{GKG21}
\beqa
\label{4.15}
\frac{\Delta \theta_0}{\Delta \theta_{1,0}}&\in&\left(0,\frac{\mathcal L_{12}}{\lambda_{-}-\mathcal L_{11}}\right)\quad \text{if} \quad
\frac{\mathcal L_{12}}{\lambda_{-}-\mathcal L_{11}}>0,\nonumber\\
\frac{\Delta \theta_0}{\Delta \theta_{1,0}}&\in&\left(\frac{\mathcal L_{12}}{\lambda_{-}-\mathcal L_{11}},0\right)\quad \text{if} \quad
\frac{\mathcal L_{12}}{\lambda_{-}-\mathcal L_{11}}<0.\nonumber\\
\eeqa

\begin{figure}[h]
	\centering
	\includegraphics[width=0.9\columnwidth]{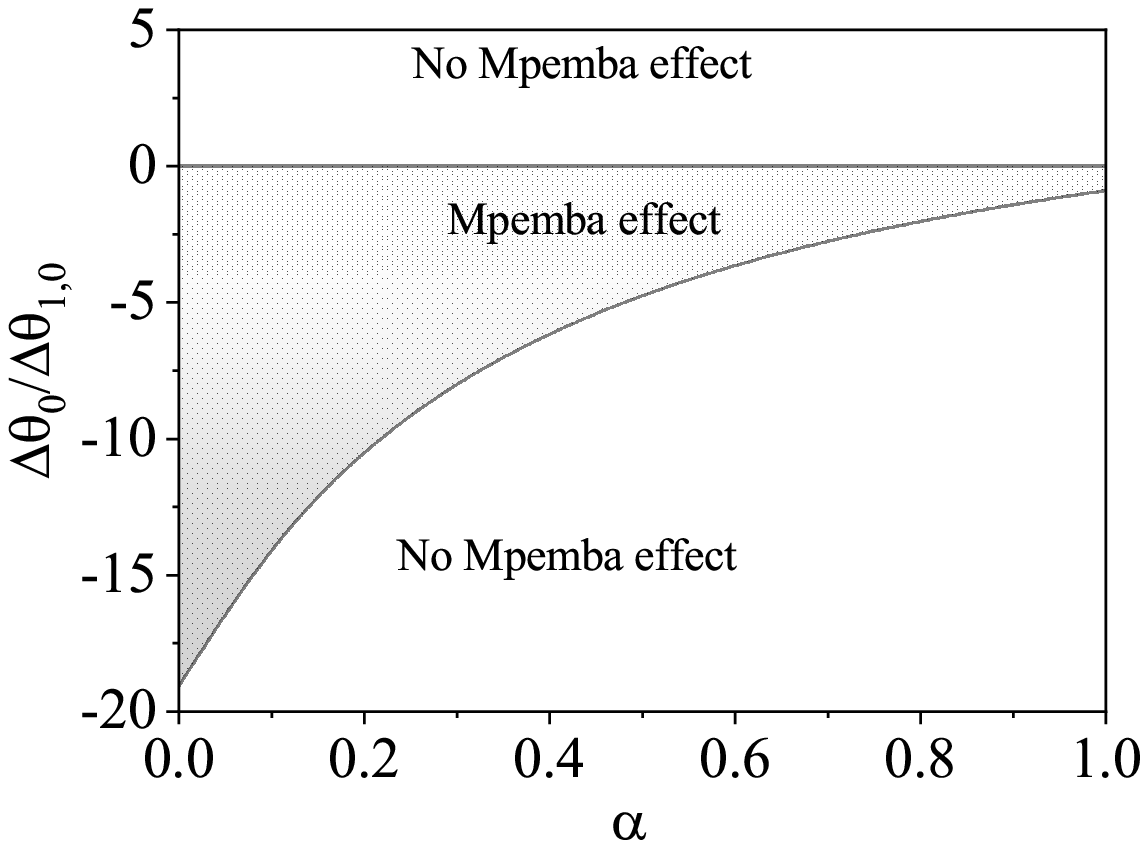}\vspace{5mm}\\
	\includegraphics[width=0.9\columnwidth]{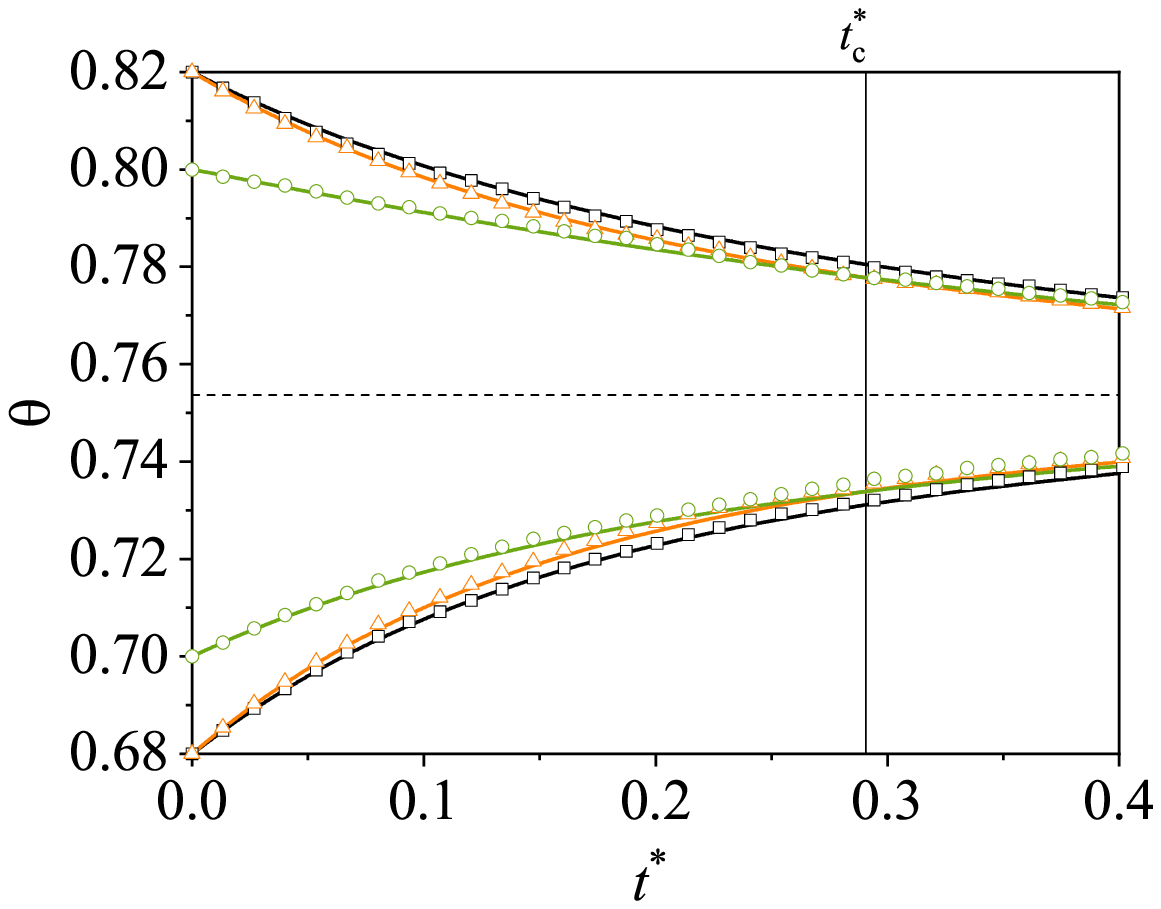}
	\caption{Top panel: phase diagram of the necessary initial condition $\Delta \theta_0/\Delta \theta_{1,0}$ as a function of the common coefficient of restitution $\alpha$. Bottom panel: relaxation of the (reduced) temperature $\theta$ towards the steady state for $\alpha=0.8$. Solid lines represent theoretical results and symbols DSMC data. The initial conditions for the temperature difference ratio $\Delta \theta_0/\Delta \theta_{1,0}$ are: $-1$ (green and orange lines and symbols) and $-4$ (green and black lines and symbols). The theoretical value of $t^*_\text{c}$ is also plotted with a vertical line. The remaining parameters in both panels are $d=3$, $m_1/m_2=5$, $\sigma_1/\sigma_2=1$, $x_1=\frac{1}{2}$, $T_{\text{ex}}^*=1$, and $\phi=0.1$. The dashed horizontal line represents the steady value $\theta_\text{s}$.}
	\label{fig3}
\end{figure}

A phase diagram showing the necessary conditions appearing in Eq.\ \eqref{4.15} as a function of the common coefficient of restitution $\alpha$ is plotted in the top panel of Fig.\ \ref{fig3}. We consider here an equimolar mixture ($x_1=\frac{1}{2}$) of hard spheres ($d=3$) of equal diameters ($\sigma_1=\sigma_2$) but different masses ($m_1=5m_2$) at moderate densities ($\phi=0.1$).  As expected, the inelasticity of collisions enlarges the region where the initial conditions lead to a crossover in the temperature relaxations. From a kinetic point of view, as inelasticity grows, particles of the hotter sample A suffer more collisions per time so, the loss of energy is emphasized when compared with the colder sample B. Thus, the inelasticity brings the relaxation curves of the two samples together and increases the possibilities of the occurrence of the Mpemba-like effect. However, the influence of the cooling rate in the time evolution of temperatures must be analyzed in conjunction with the action of the interstitial fluid. As already pointed out in Ref.\ \onlinecite{GKG21}, in the case of small inelasticity (values of $\alpha$ close to 1), the influence of the cooling rate in the relative behavior of the two samples can be neglected since it generally represents less than 10\% of the external fluid impact. On the other hand, at moderate inelasticity, the origin of the Mpemba-like effect falls on the heterogeneity of the coefficients $\lambda_i$. This discrimination on the way of transfer energy from the bath to the components of the mixture causes uneven decays of the partial temperatures towards the steady state. Hence, since the global temperature is a sum of the partial temperatures weighted by their respective mole fractions, we select the partial temperature of the component whose interaction with the bath is more effective to be the further one from the steady state. In this way, the relaxation time of the hotter sample can be reduced. In the specific case of Fig.\ \ref{fig3}, we consider a mixture of two components identical in every way except for their masses ($m_1/m_2>1$). Due to inertial effects, the transmission of momentum (and hence the transmission of kinetic energy) between the interstitial fluid and the lighter component is smoother. That is the reason why the necessary initial temperature difference $\theta_\text{A}^{(0)}-\theta_\text{B}^{(0)}=T^{(0)}_{1,\text{A}}/T^{(0)}_{2,\text{A}}-T^{(0)}_{1,\text{B}}/T^{(0)}_{2,\text{B}}<0$. On the other hand, as inelasticity increases, the action of the cooling rate becomes more relevant and a competition of both mechanisms arises.

Once discussed the constraint in the initial conditions needed for the crossover to happen, we analyze the fulfillment of Eq.\ \eqref{4.15}. To this purpose, a cooling and a heating transition towards the steady state is illustrated in the bottom panel of Fig.\ \ref{fig3}. Here, we assume the same mechanical conditions as in the phase diagram but we pick up a value for the coefficient of restitution ($\alpha=0.8$). According to Eq.\ \eqref{4.15}, the initial conditions $\Delta \theta_0/\Delta \theta_{1,0}$ must be in the range comprised between $\mathcal L_{12}/(\lambda_{-}-\mathcal L_{11})\simeq -2$ and 0. For this reason, we choose one of the initial conditions to belong to this interval ($\Delta \theta_0/\Delta \theta_{1,0}=-1$) and the other to be outside this interval ($\Delta \theta_0/\Delta \theta_{1,0}=-4$). Specific details of the initial conditions used in the above panels can be found in Table \ref{table1}.  The solid lines are the theoretical results as derived from the Enskog equation \eqref{4.1} and symbols refer to the results obtained via DSMC simulations. It is clearly shown the reliability  of conditions \eqref{4.15} and an excellent agreement between theory and simulations. Moreover, it is worth noting also the accuracy of the expression \eqref{4.14} for the crossing time $t^*_\text{c}$.

\begin{table}[h]
	\centering
	\begin{tabular}{|c|c|c|c|c|}
		\hline
		  & $\theta_0$      & $\theta_{1,0}$   & $\theta_0$      & $\theta_{1,0}$     \\ \hline
	Color	&\multicolumn{2}{|c|}{Cooling cases}   &         \multicolumn{2}{|c|}{Heating cases}                                                                                                \\ \hline
		Black                        & 0.82         & 0.815                         & 0.68        & 0.685           \\ \hline
		Orange                       & 0.82         & 0.8   
	                     & 0.68        & 0.7          \\ \hline
		Green                      & 0.8         & 0.82                       & 0.7        & 0.68              \\ \hline
	\end{tabular}
	\caption{Initial values of the (reduced) temperatures $\theta_0$ and partial temperatures $\theta_{1,0}$ used to generate the relaxation curves shown in the right panel of Fig.\ \ref{fig3}.}
	\label{table1}
\end{table}

The linearization of the Enskog equation has allowed us to give a simple explanation of the different mechanisms involved in the ocurrence of the Mpemba-like effect. There are, however, situations where small deviations from the steady state cannot be assumed. In this case, no explicit expressions for $\Delta \theta_0/\Delta \theta_{1,0}$ and $t_\text{c}^*$ can be achieved. These scenarios include the so-called \emph{large} and \emph{non-monotonic} Mpemba effects. The latter refers to crossovers in the temperature relaxation when at least one temperature presents non-monotonic evolution. In the present work, we follow similar steps as those previously made in Ref.\ \onlinecite{GKG21} to establish the necessary but not sufficient conditions for the emergence of these out-from-equilibrium phenomena.

Let us consider again two identical samples A and B whose initial temperatures $\theta_{\text{A}}^{(0)}$ and $\theta_{\text{B}}^{(0)}$ and/or partial temperatures $\theta_{1,\text{A}}^{(0)}$ and $\theta_{1,\text{B}}^{(0)}$ are significantly far away from the steady state. At the initial stages of the evolution, the condition needed for a crossover in the temperatures evolution relies on the relative behavior of the initial slopes $\Phi(\theta_{\text{A}}^{(0)},\theta_{1,\text{A}}^{(0)})$ and $\Phi(\theta_{\text{B}}^{(0)},\theta_{1,\text{B}}^{(0)})$. If we assume the sample A to be hotter than B, we must choose $\Phi_\text{A}<\Phi_\text{B}$ at the initial stages of the evolution to observe the occurrence of the Mpemba effect. Next step is to analyze the dependence of the function $\Phi$ on $\theta_1$. By doing so, we can establish some criterion for the selection of the initial partial temperature $\theta_{1}^{(0)}$ as a function of $\theta^{(0)}$. For the sake of simplicity, as proven in Ref.\ \onlinecite{GKG21}, we assume first that the influence of the inelasticity in collisions on the relative behavior of the evolution of temperatures is negligible as compared with the action of the bath. Thus, we perform the derivative of $\Phi_3$ with respect to $\theta_1$ at fixed $\theta$. The result is
\beq\label{4.16}
\left(\frac{\partial\Phi_3}{\partial\theta_1}\right)_{\theta}=2x_1(\lambda_2-\lambda_1)
\eeq
which is always a positive (negative) function if $\lambda_2>\lambda_1$ ($\lambda_2<\lambda_1$). Therefore, keeping in mind that $\theta_\text{A}^{(0)}>\theta_\text{B}^{(0)}$, then
\beqa\label{4.17}
\frac{\Delta \theta_0}{\Delta \theta_{1,0}}>0\quad&\text{if}&\quad \lambda_1>\lambda_2,\nonumber\\
\frac{\Delta \theta_0}{\Delta \theta_{1,0}}<0\quad&\text{if}&\quad \lambda_1<\lambda_2
\eeqa
are the required conditions for the presence of the Mpemba effect. Unlike the linear case, the fulfillment of Eq.\ \eqref{4.17} do not constraint the region that the initial conditions must belong to. So, in order to achieve the crossover, the difference between the initial slopes must be selected to be large enough.

\begin{figure}[h]
	\centering
	\includegraphics[width=0.9\columnwidth]{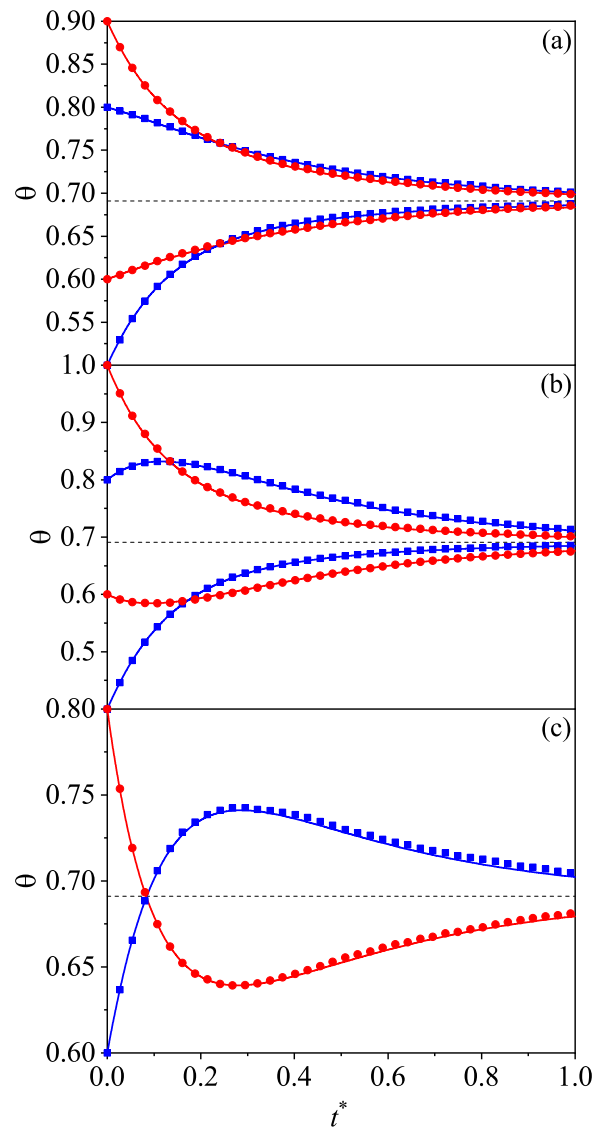}
	\caption{Relaxation of the (reduced) temperature $\theta$ towards the steady state for $\alpha=0.8$, $d=3$, $m_1/m_2=5$, $\sigma_1/\sigma_2=1$, $x_1=\frac{1}{2}$, $T_{\text{ex}}^*=1$, and $\phi=0.1$. Solid lines represent theoretical results and symbols DSMC data. (a) Large Mpemba effect: the initial conditions for the temperature difference ratio are $\Delta \theta_0/\Delta \theta_{1,0}=-1$ in both the heating and the cooling transitions. (b)Non-monotonic Mpemba effect: the initial conditions for the temperature difference ratio $\Delta \theta_0/\Delta \theta_{1,0}$ are $-2/3$ in the heating process and $-1/2$ in the cooling transition. (c) Mixed Mpemba effect: the initial condition for the temperature difference ratio is $\Delta \theta_0/\Delta \theta_{1,0}=-2/7$. The dashed horizontal lines represent the steady value $\theta_\text{s}$.}
	\label{fig3.2}
\end{figure}

Examples of the large and non-monotonic Mpemba effects are plotted in Fig.\ \ref{fig3.2} for the same parameters as in Fig.\ \ref{fig3} except for the common coefficient of restitution ($\alpha=0.7$). Since $\lambda_1<\lambda_2$, the initial temperature ratio is chosen so that $\Delta \theta_0/\Delta \theta_{1,0}<0$ (more details can be found in Table \ref{table2}). Solid lines refer to the theoretical results while symbols represent DSMC data. In Fig.\ \ref{fig3.2}(a), we observe a large Mpemba effect even when the initial temperature difference is of the same order than the temperatures themselves. In comparison with the elastic case, the inelasticity enables the choice of the partial temperature to be closer for the global temperature. This fact enhances the probability to see the non-monotonic Mpemba effect because a crossover will still be possible when the partial temperature is far away from the global temperature; inducing the appearance of non-linear effects. This latter effect is showed in Figs.\ \ref{fig3.2}(b)--\ref{fig3.2}(c). On the one hand, the non-monotonic Mpemba and its inverse effect can be observed in Fig.\ \ref{fig3.2}(b). In this case, the emergence of this surprising effect is just a matter of the choice of the initial temperature $\theta_{1,0}$. On the other hand, the mixed effect, namely when one initial temperature is above and the other below the steady one (dashed horizontal line), is plotted in Fig.\ \ref{fig3.2}(c). A good agreement between the Enskog theory and simulations can be found in all the relaxation cases ensuring the use of the Maxwellian approximation to model the distribution functions in highly non-linear situations.

\begin{table}[h]
	\centering
	\begin{tabular}{|c|c|c|c|c|c|c|}
		\hline	&	\multicolumn{2}{|c|}{Figure 4(a)}    	&	\multicolumn{2}{|c|}{Figure 4(b)}     	&	\multicolumn{2}{|c|}{Figure 4(c)}   \\
		\hline
		Color  & $\theta_0$      & $\theta_{1,0}$   & $\theta_0$      & $\theta_{1,0}$ & $\theta_0$      & $\theta_{1,0}$      \\ \hline
		\multicolumn{7}{|c|}{Cooling cases}                                                                                        \\ \hline
		Red                        & 0.9         & 0.8    & 1.0         & 0.8    & 0.8         & 0.3       \\ \hline
		Blue                      & 0.8         & 0.9   & 0.8         & 1.2   &\multicolumn{2}{c|}{---}    \\ \hline
		
		\multicolumn{7}{|c|}{Heating cases}                                                                                        \\ \hline
		Red                       & 0.6        & 0.5       & 0.6        & 0.3    &\multicolumn{2}{c|}{---}          \\ \hline
		Blue                     & 0.5       & 0.6  & 0.4        & 0.6      & 0.6         & 1.0             \\ \hline
		
	\end{tabular}
	\caption{Initial values of the (reduced) temperatures $\theta_0$ and partial temperatures $\theta_{1,0}$ used to generate the relaxation curves shown in  Fig.\ \ref{fig3.2}.}
	\label{table2}
\end{table}


\section{Steady state. Comparison between theory and DSMC simulations}
\label{sec5}

As we have discussed in section \ref{sec3}, the system achieves a \emph{steady} state for sufficiently long times. The stationary state was widely studied years ago in Ref.\ \onlinecite{KG14} where the reliability of the approximate solution to the set of Enskog equations for the temperature ratio $T_1/T_2$ and the cumulants $c_1$ and $c_2$ for a granular mixture driven by a stochastic bath with friction was assessed by molecular dynamics simulations over a wide range of the parameter space. The comparison shows a good agreement for the temperature ratio between theory and simulations for dilute and moderate densities. This good agreement contrasts with the comparison performed in \emph{dry} (or undriven) granular mixtures \cite{DHGD02} where important differences between theory and molecular dynamics simulations for $T_1/T_2$ were found for moderately dense mixtures. Regarding the comparison carried out in Ref.\ \onlinecite{KG14} for the cumulants, the results show a good agreement for dilute driven mixtures but systematic significant deviations appear as the density increases. Given that molecular dynamics avoids any assumption inherent in the kinetic theory (such as molecular chaos hypothesis), it is not clear whether the origin of the differences between theory and molecular dynamics simulations are due to the failure of the Enskog kinetic theory at high densities and/or strong inelasticity or the approximations made in solving the Enskog kinetic equation. To clarify this point, we compare in this section the (approximate) Enskog results for $T_1/T_2$, $c_1$, and $c_2$ with those obtained by numerically solving the Boltzmann--Enskog equation by means of the DSMC method. \cite{B94} Since the DSMC method (which is also based on the molecular chaos assumption) has been proved to be a powerful tool for numerically solving the Boltzmann--Enskog equation, it is quite apparent that the present comparison allow us to gauge the degree of accuracy of the approximations involved in the determination of the temperature ratio and the cumulants.

\begin{figure}[h]
	\centering
	\includegraphics[width=0.9\columnwidth]{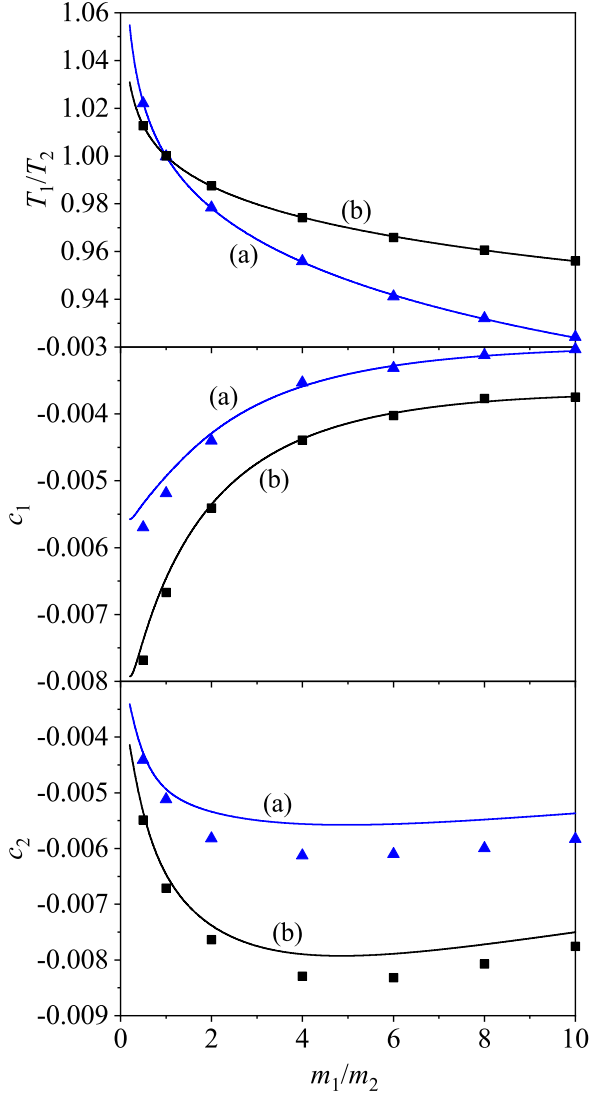}
	\caption{Case I: Plot of the temperature ratio $T_1/T_2$ and the cumulants $c_1$
		and $c_2$ as a function of the mass ratio $m_1/m_2$ for $\sigma_1/\sigma_2 = \phi_1/\phi_2 = 1$, and two
		different values of the (common) coefficient of restitution $\alpha$: $\alpha = 0.8$ (a) (blue
		lines and triangles) and $\alpha = 0.9$ (b) (black lines and squares). The lines are the
		Enskog predictions and the symbols refer to the DSMC simulation results. The remaining parameters are $T_\text{ex}^*=1$, $\phi=0.1$, and $d=3$.}
	\label{fig4}
\end{figure}
\begin{figure}[h]
	\centering
	\includegraphics[width=0.9\columnwidth]{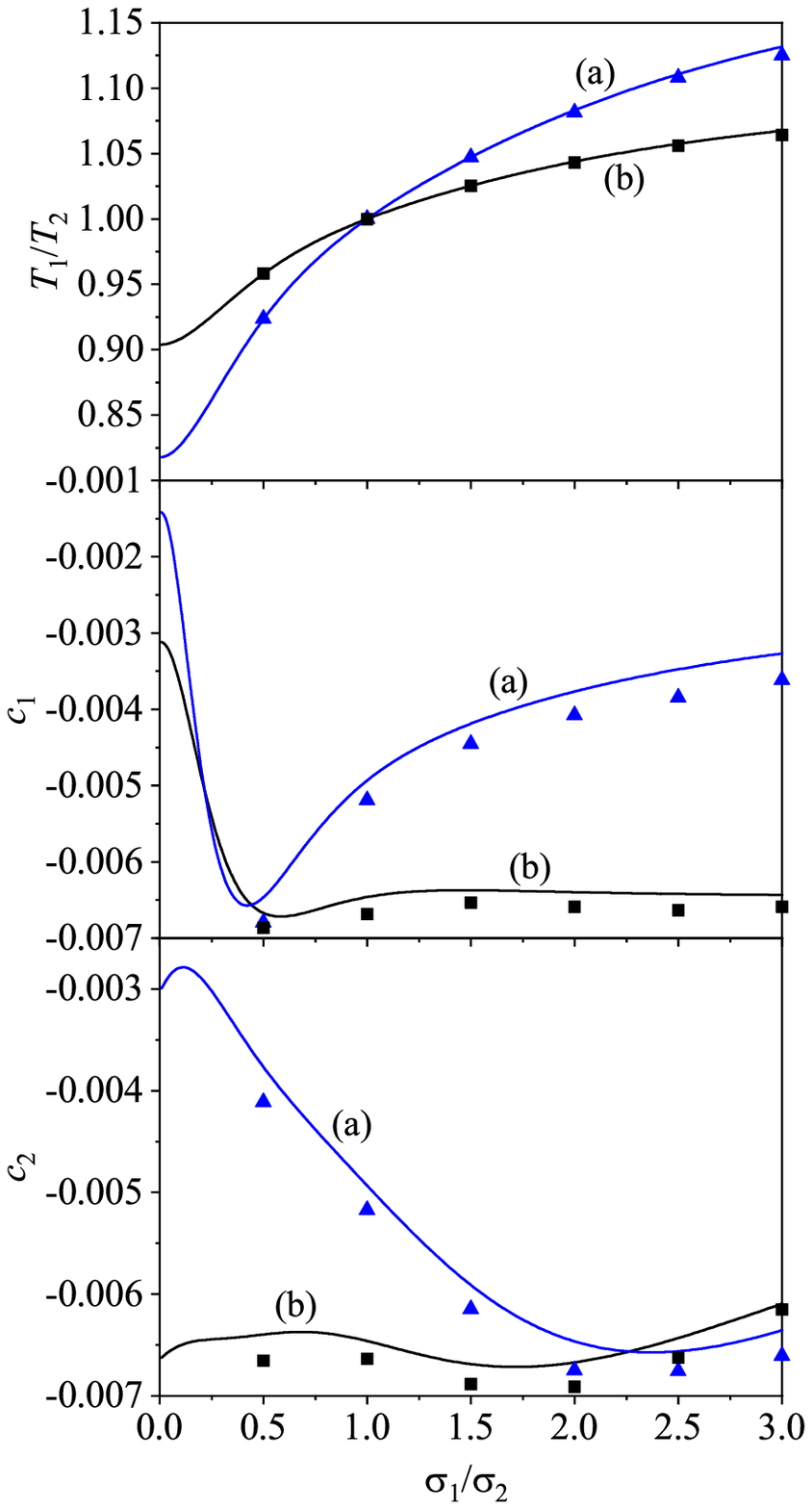}
	\caption{Case II: Plot of the temperature ratio $T_1/T_2$ and the cumulants $c_1$
		and $c_2$ as a function of the size ratio $\sigma_1/\sigma_2$ for $m_1/m_2 =\phi_1/\phi_2 = 1$, and two
		different values of the (common) coefficient of restitution $\alpha$: $\alpha = 0.8$ (a) (blue
		lines and triangles) and $\alpha = 0.9$ (b) (black lines and squares). The lines are the
		Enskog predictions and the symbols refer to the DSMC simulation results. The remaining parameters are $T_\text{ex}^*=1$, $\phi=0.1$, and $d=3$.}
	\label{fig5}
\end{figure}
\begin{figure}[h]
	\centering
	\includegraphics[width=0.9\columnwidth]{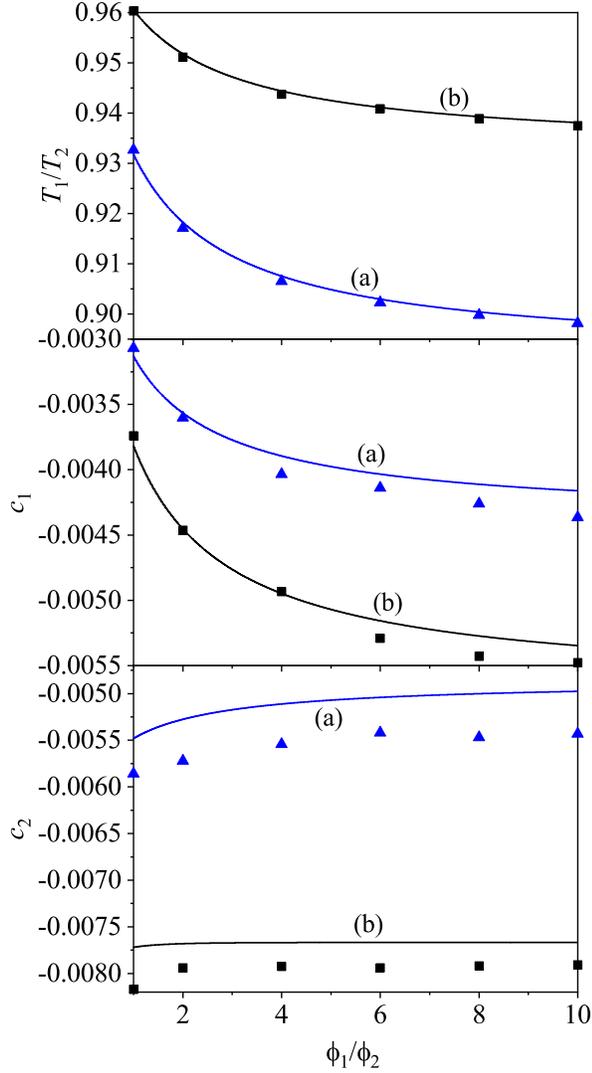}
	\caption{Case III: Plot of the temperature ratio $T_1/T_2$ and the cumulants $c_1$
		and $c_2$ as a function of the partial density ratio $\phi_1/\phi_2$ for $m_1/m_2 =8$, $\sigma_1/\sigma_2 = 1$, and two
		different values of the (common) coefficient of restitution $\alpha$: $\alpha = 0.8$ (a) (blue
		lines and triangles) and $\alpha = 0.9$ (b) (black lines and squares). The lines are the
		Enskog predictions and the symbols refer to the DSMC simulation results. The remaining parameters are $T_\text{ex}^*=1$, $\phi=0.1$, and $d=3$.}
	\label{fig6}
\end{figure}

The steady state is defined by the conditions $\partial_\theta \theta_1=\partial_\theta \theta_2=0$. According to Eq.\ \eqref{3.12}, the above conditions imply that $\Lambda_1=\Lambda_2=0$, which yields the following set of equations for the partial temperatures $\theta_1$ and $\theta_2$:
\beq
\label{5.1}
2\gamma_1^*\left(1-\theta_1\right)=\theta_1\left(\zeta_{10}+\zeta_{11}c_1+\zeta_{12}c_2\right),
\eeq
\beq
\label{5.2}
2\gamma_2^*\left(1-\theta_2\right)=\theta_2\left(\zeta_{20}+\zeta_{21}c_1+\zeta_{22}c_2\right).
\eeq
Upon writing Eqs.\ \eqref{5.1}--\eqref{5.2} use has been made of the expansions \eqref{3.17}. In Eqs.\ \eqref{5.1} and \eqref{5.2} and the remaining part of this section, it is understood that all the quantities are evaluated in the steady state. Equations \eqref{5.1} and \eqref{5.2} are coupled to those of the cumulants $c_1$ and $c_2$. The equations for the cumulants are obtained from Eq.\ \eqref{3.15} by taking the steady-state conditions $\Lambda_1=\Lambda_2=0$. This leads to the following set of algebraic linear equations:
\beqa
\label{5.3}
& &d(d+2)\Big(\frac{\theta_1}{M_1\theta}\Big)^2\gamma_1^*(1-\theta_1^{-1})-\Sigma_{10}=\nonumber\\ & &\Big[\Sigma_{11}-d(d+2)\Big(\frac{\theta_1}{M_1\theta}\Big)^2
\gamma_1^*\Big]c_1+\Sigma_{12}c_2,
\eeqa
\beqa
\label{5.4}
& &d(d+2)\Big(\frac{\theta_2}{M_2\theta}\Big)^2\gamma_2^*(1-\theta_2^{-1})-\Sigma_{20}=\nonumber\\ & &\Big[\Sigma_{22}-d(d+2)\Big(\frac{\theta_2}{M_2\theta}\Big)^2
\gamma_2^*\Big]c_2+\Sigma_{21}c_1,
\eeqa
where use has been made of the expansion \eqref{3.18}.

Solution to the set of equations \eqref{5.1}--\eqref{5.4} provides the stationary values of the ratio of partial temperatures $T_1/T_2$ and the cumulants $c_1$ and $c_2$. These quantities are given as a functions of the dimensionality $d$, the (reduced) background temperature $T_\text{ex}^*$, the mass ratio $m_1/m_2$, the concentration ratio $\phi_1/\phi_2$, the ratio of diameters $\sigma_1/\sigma_2$, the density $\phi$, and the coefficients of restitution $\al_{11}$, $\al_{22}$, and $\al_{12}$. Since the parameter space of the problem is large, as usual and to reduce the number of independent parameters, we consider a three-dimensional system ($d=3$), a (reduced) bath temperature $T_\text{ex}^*=1$, a moderate density $\phi=0.1$, and a common coefficient of restitution $\al\equiv \al_{11}=\al_{22}=\al_{12}$. This reduces the parameter space to four dimensionless quantities: $\left\{m_1/m_2, \phi_1/\phi_2, \sigma_1/\sigma_2, \al\right\}$.

As in Ref.\ \onlinecite{KG14}, the set of dimensionless quantities $\Xi\equiv \left\{T_1/T_2, c_1, c_2\right\}$ have been obtained from the approximate theory and DSMC simulations in three different cases. Two different values of $\al$ have been considered in each case: $\alpha=0.9$ (moderate inelasticity) and $\al=0.8$ (strong inelasticity). In the first case (case I) the set $\Xi$ is determined as a function of the mass ratio $m_1/m_2$ for $\phi_1/\phi_2=\sigma_1/\sigma_2=1$, while in the second case (case II) $\Xi$ is obtained as a function of the ratio of diameters $\sigma_1/\sigma_2$ for $m_1/m_2=\phi_1/\phi_2=1$. Finally, in case III, $\Xi$ is given as a function of concentration $\phi_1/\phi_2$ for $m_1/m_2=8$ and $\sigma_1/\sigma_2=1$. Given the disparity of parameters of the mixture analyzed in the three different cases, the test of the approximate kinetic theory can be considered as stringent.

Case I is shown in Fig.\ \ref{fig4}. While the solid lines correspond to the (approximate) theoretical results, the symbols represent the Monte Carlo simulation data (squares for $\al=0.9$ and triangles for $\al=0.8$). As expected, the extent of the energy nonequipartition increases with the mass disparity of the mixture. On the other hand, the departure form energy equipartition is more noticeable in dry granular mixtures than in binary granular suspensions. Figure \ref{fig4} highlights the excellent agreement between theory and simulations for the temperature ratio, even for quite disparate masses. With respect to the cumulants, we observe that the magnitude of $c_1$ and $c_2$ is much smaller than that of a dry granular mixture. \cite{GD99b,MG02,DHGD02} In addition, while the theoretical results for $c_1$ compare well with simulations, some discrepancies are found in the case of the cumulant $c_2$ (specially for large values of the mass ratio) since the theory slightly underestimates the value of $c_2$. In any case, the quantitative discrepancies between theory and simulations are of the same order as those observed in the dry granular limit case \cite{MG02} since the largest relative error of $c_2$ is about 9\%.

Figure \ref{fig5} shows the results of case II, $\Xi$ as function of the ratio of diameters $\sigma_1/\sigma_2$. As in Fig.\ \ref{fig4}, the agreement is again excellent for the temperature ratio; more significant discrepancies are observed for both cumulants in case II than in case I. These discrepancies could be likely mitigated by considering nonlinear terms in $c_1$ and $c_2$ in the approximate theory and/or by considering more terms in the Sonine polynomial expansion of $\varphi_i(\mathbf{c})$. However, given that the price to be paid for considering these type of terms is very high (since the involved calculations would be very cumbersome), we think that the approximate theory reported here is still an accurate approach to estimate the cumulants. In fact, as in case I, the largest relative error found in Fig.\ \ref{fig5} for $c_1$ is 9.4 \% and 8.2 \% for $c_2$. Finally, Fig.\ \ref{fig6} show case III, $\Xi$ versus the concentration $\phi_1/\phi_2$. It is quite apparent that Fig.\ \ref{fig6} exhibits similar trends as those observed before for Figs.\ \ref{fig4} and \ref{fig5}: while $T_1/T_2$ displays an excellent agreement between theory and simulations, there are small differences for the cumulants.

\section{Linear stability analysis of the steady solution}
\label{sec6}

\begin{figure}[h]
	\centering
	\includegraphics[width=0.9\columnwidth]{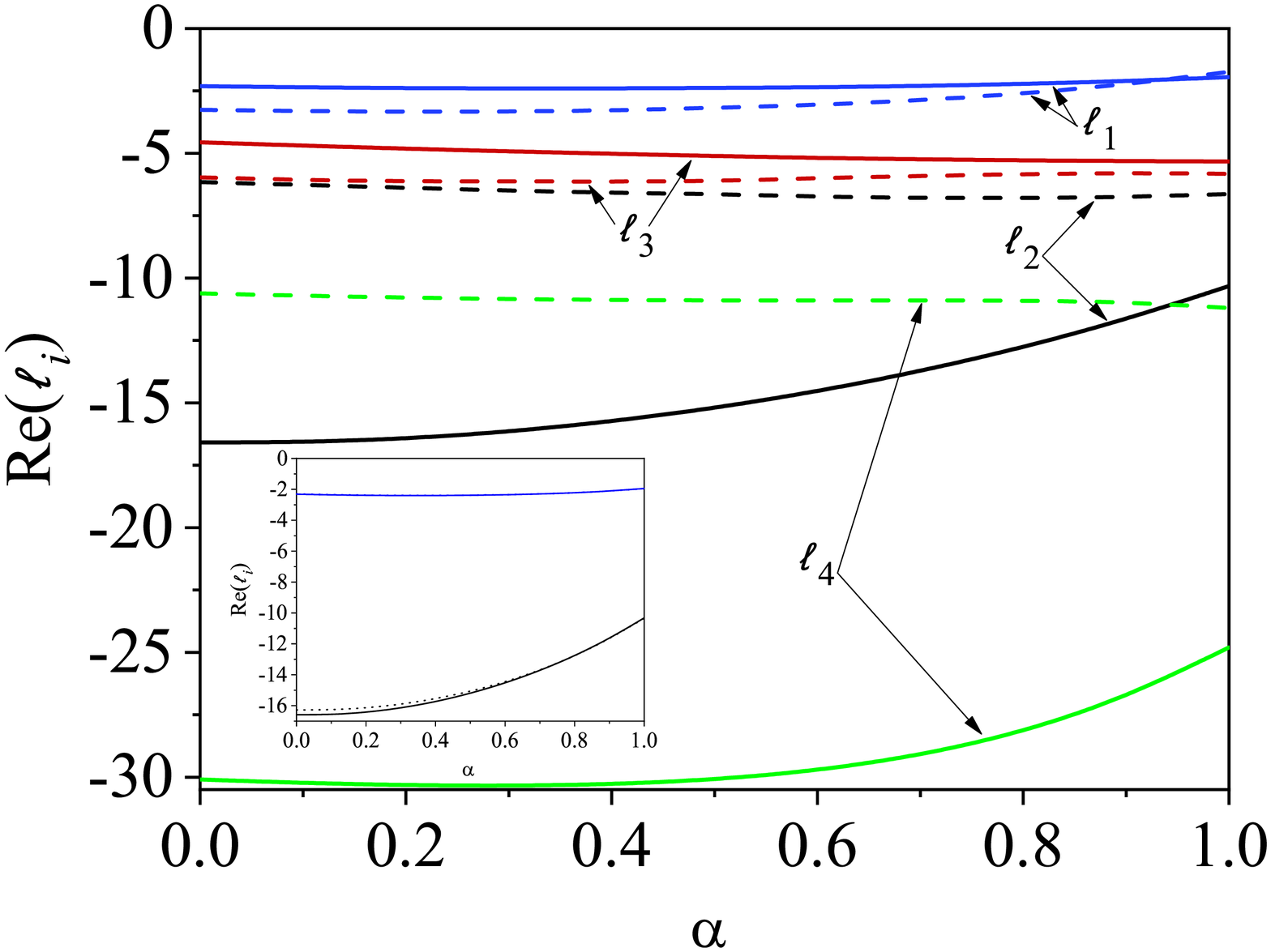}
	\caption{ Plot of the real parts of the eigenvalues $\ell_i$	($i = 1, 2, 3, 4$) of the matrix $\mathcal{L}$ for $\sigma_1/\sigma_2=1$, $x_1=\frac{1}{2}$, $T_\text{ex}^*=1$, $\phi=0.1$, and $d=3$. Solid lines corresponds to $m_1/m_2=10$ and dashed lines to $m_1/m_2=\frac{1}{2}$. The inset shows a comparison of $\ell_i$ ($i=1,2$) for $m_1/m_2=10$ when cumulants (solid lines) and no cumulants (dashed lines) are considered.}
	\label{fig7}
\end{figure}

Although the study offered in section \ref{sec5} has been focused on the determination of the temperature ratio and the cumulants in steady state conditions, it is worthwhile to analyze if the stationary solution actually conforms indeed a linearly stable solution. To study the stability of this steady solution we will take into account the effect of cumulants $c_1$ and $c_2$ on the evolution equations of $\theta$ and $\theta_1$. Retaining only linear terms in the above cumulants, the evolution equations of $\theta$, $\theta_1$, $c_1$, and $c_2$ can be written, respectively, as
\begin{widetext}
\beqa
\label{6.1}
\frac{\partial \theta}{\partial t^*}&=&2\Big[x_1\lambda_1+x_2\lambda_2-x_1\left(\lambda_1-\lambda_2\right)\theta_1-\lambda_2\theta\Big]
-\theta^{1/2}\Big\{x_1\theta_1\Big[\zeta_{10}-\zeta_{20}
\nonumber\\
& & +\left(\zeta_{11}-\zeta_{21}\right)c_1+\left(\zeta_{12}-\zeta_{22}\right)c_2\Big]
+\theta\left(\zeta_{20}+\zeta_{21}c_1+\zeta_{22}c_2\right)\Big\},
\eeqa
\beq
\label{6.2}
\frac{\partial \theta_1}{\partial t^*}=2\lambda_1(1-\theta_1)-\theta^{1/2}\theta_1\left(\zeta_{10}+\zeta_{11}c_1+\zeta_{12}c_2\right),
\eeq
\beqa
\label{6.3}
\frac{\partial c_1}{\partial t^*}&=&-4\lambda_1\left(\theta_1^{-1}-1\right)(1+c_1)+2\theta^{1/2}\Big[\zeta_{10}+\left(\zeta_{10}+\zeta_{11}\right)c_1+\zeta_{12}c_2\Big]
\nonumber\\
& &+4\lambda_1\theta_1^{-1}
+\frac{4\theta^{5/2}}{d(d+2)}\Big(\frac{M_1}{\theta_1}\Big)^2\Big(\Sigma_{10}+\Sigma_{11}c_1+\Sigma_{12}c_2\Big),
\eeqa
\beqa
\label{6.4}
\frac{\partial c_2}{\partial t^*}&=&-4\lambda_2\left(\theta_2^{-1}-1\right)(1+c_2)+2\theta^{1/2}\Big[\zeta_{20}+\left(\zeta_{20}+\zeta_{22}\right)c_2+\zeta_{21}c_1\Big]
\nonumber\\
& &+4\lambda_2\theta_2^{-1}
+\frac{4\theta^{5/2}}{d(d+2)}\Big(\frac{M_2}{\theta_2}\Big)^2\Big(\Sigma_{20}+\Sigma_{21}c_1+\Sigma_{22}c_2\Big).
\eeqa
\end{widetext}
Here, we recall that $\theta_2(t^*)=x_2^{-1}\left[\theta(t^*)-x_1 \theta_1(t^*)\right]$ and use has been made of the expansions \eqref{3.17} and \eqref{3.18} for obtaining Eqs.\ \eqref{6.1}--\eqref{6.4}.

Now, as in section \ref{sec3}, one looks for solutions of the form
\begin{align}
\label{6.5}
\theta(t^*)&=\theta_\text{s}+\delta \theta(t^*), \quad &\theta_1(t^*)=& \ \theta_\text{1,s}+\delta \theta_1(t^*),\nonumber\\   c_1(t^*)&=c_{1,\text{s}}+\delta c_1(t^*), \quad &c_2(t^*)=& \ c_{2,\text{s}}+\delta c_2(t^*),
\end{align}
and neglects nonlinear terms in the perturbations $\left\{\delta \theta, \delta \theta_1, \delta c_1, \delta c_2\right\}$. If the real parts of the eigenvalues $\ell_i$ ($i=1,2,3,4$) are negative, the steady solution $\left\{\theta_\text{s}, \theta_{1,\text{s}}, c_{1,\text{s}}, c_{2,\text{s}}\right\}$ is (linearly) stable. The expressions of the eigenvalues $\ell_i$ are very large and will be omitted here for the sake of brevity. On the other hand, in the simplest case where the cumulants $c_i$ are neglected in the determination of the cooling rates $\zeta_i^*$ and the collisional moments $\Sigma_{i}$, the time evolution of  $\delta \theta(t^*)$ and $\delta \theta_1(t^*)$ is governed by the eigenvalues $\ell_1$ and $\ell_2$ defined by Eq.\ \eqref{4.13}. A careful analysis of the eigenvalues shows that $\text{Re}(\ell_i)<0$ $(i=1,2,3,4)$, so that the steady state is always stable.

As an illustration, the real parts of the eigenvalues $\ell_i$ are plotted in Fig.\ \ref{fig7} against the common coefficient of restitution $\alpha\equiv\alpha_{12}=\alpha_{12}=\alpha_{22}$ for $d=3$, $\sigma_1/\sigma_2=1$, $x_1=\frac{1}{2}$, $\phi=0.1$, and $T_\text{ex}^*=1$. Two values of the mass ratio are considered: $m_1/m_2=10$ and $m_1/m_2=\frac{1}{2}$. It is quite
apparent that the real part of the eigenvalues $\ell_i$ is always negative. Moreover, the inset of figure \ref{fig7} shows a comparison of the eigenvalues $\ell_1$ and $\ell_2$ when the cumulants are neglected versus those obtained by solving the set of coupled equations \eqref{6.1}--\eqref{6.4}. No significant discrepancies between both approaches are found in the qualitative behavior of $\ell_i$ and, hence, the reliability of the Maxwellian approximation is ensured once again.

\section{Discussion}
\label{sec7}

In this paper, we have analyzed the time-dependent homogeneous state of a binary granular suspension. The starting point of the study has been the set of two coupled Enskog kinetic equations for the velocity distribution functions $f_i(\mathbf{v};t)$ ($i=1,2$) of the solid particles. As usual, the influence of the surrounding viscous gas on the dynamics of grains has been accounted for in an effective way by means of a force constituted by two terms: a deterministic viscous drag force plus a stochastic Langevin-like term. This simple suspension model is mainly based on the assumption that the interstitial fluid is not perturbed by the grains and so, it can be considered as a thermostat at the (known) temperature $T_\text{ex}$. On the other hand, since the model is inspired in numerical and experimental results, \cite{YS09b} the friction coefficients $\gamma_i$ display a complex dependence on the partial $\phi_i$ and global $\phi=\phi_1+\phi_2$ volume fractions, and the masses $m_i$ and diameters $\sigma_i$ of the mixture [see Eqs.\ \eqref{2.7} and \eqref{2.8}].

The objective of the paper is twofold. First, we want to characterize the temporal evolution of the system towards the asymptotic steady state. In particular, we have investigated the existence of an unsteady ``hydrodynamic'' stage (where the velocity distributions $f_i$ depend on time only through the global temperature $T(t)$) before achieving the stationary regime. The existence of the above time-dependent state is crucial for deriving the corresponding Navier--Stokes hydrodynamic equations since this state plays the role of ``reference'' state in the application of the Chapman--Enskog method \cite{CC70} to granular suspensions. \cite{GKG20} As a complement to this study, we have also explored the occurrence of the so-called Mpemba-like effect (an initially hotter gas cools sooner than the colder one) in bidisperse granular suspensions. This analysis extends to inelastic collisions previous results derived for molecular suspensions. \cite{GKG21,GG21} Beyond the transient regime and as a second objective, we have also determined the temperature ratio $T_1/T_2$ and the cumulants $c_1$ and $c_2$ (which measure the departure of the distributions $f_i$ from their Maxwellian forms) in the stationary state as functions of the mass and size ratios, the concentration, the volume fraction, the coefficients of restitution, and the background temperature. It is worthwhile remarking that the (approximate) theoretical results obtained in each one of the different issues covered along the paper have been tested against DSMC simulations \cite{B94} for different systems and conditions.

Regarding the transient regime, theory and simulations have clearly shown that, after a kinetic stage and before the steady state is reached, the system evolves towards a universal unsteady hydrodynamic stage that no longer depends on the initial conditions. As for driven granular gases, \cite{GMT12,ChVG13} the distributions $f_i(\mathbf{v};t)$ have the form \eqref{3.6} where the time dependence of the scaled distributions $\varphi_i$ not only occurs through the dimensionless velocity $\mathbf{c}(t)=\mathbf{v}/v_0(t)$ but also through the scaled temperature $\theta(t)=T(t)/T_\text{ex}$. A consequence of this scaling is that the velocity moments $\theta_i(t^*)=T_i(t^*)/T_\text{ex}$ and $c_i(t^*)$ tend towards the universal functions $\theta_i(\theta(t^*))$ and $c_i(\theta(t^*))$, respectively, where the functions $\theta_i(\theta)$ and $c_i(\theta)$ are independent of the initial conditions.

With respect to the Mpemba-like effect, as expected this phenomenon is also present when collisions in the binary mixture are inelastic. However, in contrast to the analysis performed in Refs.\ \onlinecite{GKG21} and \onlinecite{GG21} for elastic collisions, the presence of the cooling term $\zeta^*$ [which gives rise to the granular new term $\Phi_4$ in the evolution equation of the temperature $\theta(t^*)$; see Eq.\ \eqref{4.1}] makes more difficult to find clean initial conditions for the occurrence of the Mpemba-like effect. To gain some insight, situations near the final asymptotic steady state have been considered first to get explicit expressions for the crossing time $t_\text{c}^*$. By analyzing the dependence of $t_\text{c}^*$ on the initial conditions, we have been able to study the necessary conditions for the effect to occur. Figure \ref{fig3} illustrates the dependence of the initial temperature ratio $\Delta \theta_0/\Delta \theta_{1,0}$ as a function of the common coefficient of restitution $\alpha$. As expected, inelasticity of collisions increases the possibilities to observe the Mpemba-like effect. Moreover, the necessary conditions given in Eq.\ \eqref{4.15} are tested again DSMC simulations in the right panel of Fig.\ \ref{fig3} for a cooling and a heating transition. The excellent agreement found between theory and simulations ensure the use of the Maxwellian approximation.

Once we have studied the Mpemba-like effect in situations close to the steady state, we have explored then non-linear situations. The coupling between $\theta$ and $\theta_1$ provokes the appearance of the large and non-monotonic Mpemba effects. In the former, the large Mpemba effect has been observed even when the initial temperature difference is about 10\% of the temperatures themselves. Inelasticity of collisions enlarges the necessary distance between $\theta_0$ and $\theta_{1,0}$ that leads to a crossover in the evolution of temperatures. Thus, non-linear effects arise and we can observe the non-monotonic and mixed Mpemba effects. Fig.\ \ref{fig3.2} illustrates the large, non-monotonic, and mixed effects for a given case and exhibits a good agreement between theory and simulations in the set of parameters considered. However, we have neglected the influence of inelasticity to determine the necessary conditions for the emergence of the Mpemba effect in the non-linear regime. Therefore, we could consider a \emph{dry} (no gas phase) granular mixture to easily draw conclusions about the effect of inelasticity in the appearance of such effect. We plan to carry out a more exhaustive study on the necessary conditions for the onset of the Mpemba-like effect in dry granular mixtures in the near future.

Finally, the stationary values of the temperature ratio and the fourth cumulants have been determined and compared with DSMC simulations. This study complements a previous comparison made in Ref.\ \onlinecite{KG14} between kinetic theory and molecular dynamics simulations. In this context, the comparison carried here in section \ref{sec4} can be seen as a test of the approximations involved in the computation of $T_1/T_2$ and $c_i$ but not as a test of the kinetic equation itself since the DSMC method does not avoid the inherent assumptions of kinetic theory (molecular chaos hypothesis). As Figs.\ \ref{fig4}--\ref{fig6} clearly show, theoretical results for $T_1/T_2$ agree very well with DSMC results for all the systems considered in the simulations. On the other hand, in the case of the cumulants, although theory compares qualitatively well with simulations, more quantitative discrepancies are found between both approaches (especially in the case of $c_2$). This quantitative disagreement between theory and simulations could be mitigated by the inclusion of cumulants of higher order as well as nonlinear terms in $c_1$ and $c_2$. However, based on previous results obtained for monocomponent granular gases \cite{BP06a,BP06b} on the possible lack of convergence of the Sonine polynomial expansion, the absolute value of the higher order cumulants could increase with inelasticity. In this case, the Sonine expansion could be not relevant in the sense that one would need to retain a large number of Sonine coefficients to achieve an accurate estimate of the fourth-degree cumulants.

Although the results derived in this paper have been focused on smooth inelastic spheres, the extension to inelastic \emph{rough} hard spheres is a very challenging problem. This study could allow us to assess the impact of the solid body friction on the applicability of a hydrodynamic description to granular suspensions and/or the occurrence of the Mpemba effect. Based on previous results, \cite{GG20} we expect that the effect of roughness on the dynamic properties of grains can play an important role. We will work on this issue in the near future.

In summary, we believe our results provide additional support to the validity of hydrodynamics for studying time-dependent homogeneous states in multicomponent granular suspensions. As said before, this conclusion is relevant since the local version of the time-dependent homogeneous state is considered as the zeroth-order approximation in the Chapman--Enskog expansion. In addition, we have also shown the occurrence of the Mpemba-like effect in bidisperse granular suspensions for situations close to and far away from the asymptotic stationary state. In both cases, approximate theoretical results agree very well with DSMC simulations. As a complement of the previous studies, the temperature ratio $T_1/T_2$ and the fourth-degree cumulants $c_i$ have been also determined in the stationary state. While theory shows an excellent agreement with simulations for $T_1/T_2$, some differences are found in the case of the cumulants. However, these differences are relatively small and in fact they are of the same order as those observed in the homogeneous cooling state for undriven granular mixtures. \cite{GD99b,MG02}

\acknowledgments

The work has been supported by the Spanish Government through Grant No. PID2020-112936GB-I00 and by the Junta de Extremadura (Spain) Grant Nos. IB20079 and GR18079, partially financed by ``Fondo Europeo de Desarrollo Regional'' funds. The research of R.G.G. also has been supported by the predoctoral fellowship BES-2017-079725 from the Spanish Government.
\\

 \noindent \textbf{CONFLICT OF INTEREST}\\

 The authors have no conflicts to disclose.\\

 \noindent \textbf{DATA AVAILABILITY}\\

The data that support the findings of this study are available from the corresponding author upon reasonable request.

\begin{widetext}
\appendix

\section{Expressions for the partial cooling rates and the fourth degree collisional moments}
\label{appA}

In this Appendix we display the explicit expressions of the (reduced) partial cooling rates $\zeta_i^*$ and the fourth degree collisional moments $\Sigma_i$. Their forms are provided by Eqs.\ \eqref{3.17} and \eqref{3.18} when nonlinear terms in $c_i$ are neglected. The corresponding expressions of $\zeta_{ij}$ and $\Sigma_{ij}$ are given by \cite{KG14}
\beqa
\label{a1}
\zeta_{10}&=&\frac{\sqrt{2}\pi^{(d-1)/2}}{d\Gamma\left(\frac{d}{2}\right)}x_1 \chi_{11}\left(\frac{\sigma
	_1}{\sigma_{12}}\right)^{d-1}\beta_1^{-1/2} (1-\alpha_{11}^2)+\frac{4\pi^{(d-1)/2}}{d\Gamma\left(\frac{d}{2}\right)}x_2 \chi_{12}\mu_{21} \left(\frac{1+\beta}{\beta}\right)^{1/2}(1+\alpha_{12})\beta_2^{-1/2}\nonumber\\
& & \times \left[1-\frac{1}{2}\mu_{21}(1+\alpha_{12})(1+\beta) \right],
\eeqa
\begin{eqnarray}
\label{a2}
\zeta_{11}&=&\frac{3\pi^{(d-1)/2}}{8\sqrt{2}d\Gamma\left(\frac{d}{2}\right)} x_1 \chi_{11} \left(\frac{\sigma
	_1}{\sigma_{12}}\right)^{d-1} \beta_1^{-1/2}
(1-\alpha_{11}^2)+\frac{\pi^{(d-1)/2}}{2d\Gamma\left(\frac{d}{2}\right)}x_2 \chi_{12}\mu_{21}
\frac{(1+\beta)^{-3/2}}{\beta^{1/2}}(1+\alpha_{12})\beta_2^{-1/2} \nonumber\\
& & \times
\Big[3+4\beta-\frac{3}{2}\mu_{21}(1+\alpha_{12})(1+\beta) \Big],
\end{eqnarray}
\beq
\label{a3}
\zeta_{12}=-\frac{\pi^{(d-1)/2}}{2d\Gamma\left(\frac{d}{2}\right)}x_2 \chi_{12}
\mu_{21}\left(\frac{1+\beta}{\beta}\right)^{-3/2}(1+\alpha_{12})\beta_2^{-1/2}
\Big[1+\frac{3}{2}\mu_{21}(1+\alpha_{12})(1+\beta) \Big],
\eeq
\begin{eqnarray}
\label{a4}
\Sigma_{10} &=&-\frac{\pi^{(d-1)/2}}{\sqrt{2}\Gamma\left(\frac{d}{2}\right)} \beta_{1}^{-5/2}
x_{1}\chi_{11}\left(\frac{\sigma_{1}}{{\sigma}_{12}}\right)^{d-1} \frac{3+2d+2\alpha_{11}^{2}}{2}
\left(1-\alpha_{11}^{2}\right)+\frac{2\pi^{(d-1)/2}}{\Gamma\left(\frac{d}{2}\right)} \beta_{1}^{-5/2}x_2 \chi_{12}\left( 1+\beta\right)^{-1/2}\nonumber\\
& & \times  \mu_{21}\left( 1+\alpha_{12}\right)\Big\{-\left[d+3+(d+2)\beta\right] +\frac{\mu_{21}}{2}\left(1+\alpha_{12}\right) \left( 1+\beta \right)
\left( 11+ d+\frac{d^2+5d+6}{d+3} \theta \right)\nonumber\\
& & -4\mu_{21}^{2}\left(1+\alpha_{12}\right)^{2}\left(1+\beta \right)^{2} +\mu_{21}^{3}\left(
1+\alpha_{12}\right)^{3}\left(1+\beta \right)^{3}\Big\},
\end{eqnarray}
\begin{eqnarray}
\label{a5}
\Sigma _{11} &=&-\frac{\sqrt{2}\pi^{(d-1)/2}}{\Gamma\left(\frac{d}{2}\right)} \beta_{1}^{-5/2}
x_{1}\chi_{11} \left(\frac{\sigma_{1}}{{\sigma}_{12}}\right)^{d-1} \left[\frac{d-1}{2}(1+\alpha_{11})+\frac{3}{64}
\left(10d+39+10\alpha_{11}^{2}\right) \left(1-\alpha_{11}^{2}\right)\right] \nonumber \\
& & +\frac{\pi^{(d-1)/2}}{4\Gamma\left(\frac{d}{2}\right)} \beta_{1}^{-5/2}x_2 \chi_{12}\left(1+\beta\right)^{-5/2}
\mu_{21}\left(1+\alpha_{12}\right)\Big\{-\big[ 45+15d+(114+39d)\beta+(88+32d)\beta^{2}\nonumber\\
& & +(16+8d)\beta^{3}\big]+\frac{3}{2}\mu_{21}\left(1+\alpha_{12}\right) \left(1+\beta \right) \left[ 55+5d+9(10+d)\beta+4(8+d)\beta
^{2}\right]\nonumber\\
& & -12\mu_{21}^{2}\left(1+\alpha_{12}\right)^{2}
\left(1+\beta \right) ^{2}\left(5+4\beta
\right)+15\mu_{21}^{3}\left(1+\alpha_{12}\right)^{3}\left( 1+\beta \right)^{3}\Big\},
\end{eqnarray}
\beqa
\label{a6}
\Sigma_{12}&=&\frac{\pi^{(d-1)/2}}{4\Gamma\left(\frac{d}{2}\right)}
\theta_{1}^{-5/2}x_2 \chi_{12}\beta^2\left(1+\beta\right)^{-5/2} \mu_{21}\left(
1+\alpha_{12}\right) \Big[d-1+(d+2)\beta+\frac{3}{2}\mu_{21}\left(1+\alpha_{12}\right) \left(1+\beta\right)\nonumber\\
& & \times \left[d-1+(d+2)\beta\right]
-12\mu_{21}^{2}\left(1+\alpha_{12}\right)^{2}\left(1+\beta\right)^{2}+15\mu_{21}^{3}\left(1+\alpha_{12}\right)^{3}\left(1+\beta\right)^{3}\Big] .
\eeqa
The expressions for $\zeta_{20}$, $\zeta_{21}$, $\zeta_{22}$, $\Sigma_{20}$, $\Sigma_{22}$ and $\Sigma_{21}$ can easily obtained from Eqs.\ \eqref{a1}--\eqref{a6} by changing $1 \to 2$ and $\beta\to \beta^{-1}$.
\end{widetext}


%

\end{document}